\documentclass[onecolumn,showpacs,
amsmath,amssymb,superscriptaddress,prl]{revtex4-2}

\usepackage{graphicx}
\usepackage{dcolumn}
\usepackage{bm}

\usepackage[version=3]{mhchem} 
\usepackage{amsmath}
\usepackage{esvect}

\usepackage{array}
\usepackage{graphicx}
\usepackage{graphicx}
\usepackage{dcolumn}
\usepackage{bm}
\usepackage{color}
\usepackage{pifont}
\usepackage{natbib}
\usepackage{hyperref}
\hypersetup{
colorlinks = true,
urlcolor   = blue,
linkcolor  = blue,
citecolor  = blue
}
\usepackage{nicefrac}

\usepackage{color}
\newcommand{\red}[1]{{\color{black} #1}}
\newcommand{\blue}[1]{{\color{black} #1}}

\begin{document}

\title{Pseudo-spin model of argentophilicity in honeycomb bilayered materials}

\author{Godwill Mbiti Kanyolo}
\email{gmkanyolo@mail.uec.jp; gm.kanyolo@aist.go.jp}
\affiliation{Research Institute of Electrochemical Energy (RIECEN), National Institute of Advanced Industrial Science and Technology (AIST), 1-8-31 Midorigaoka, Ikeda, Osaka 563-8577, Japan}
\affiliation{The University of Electro-Communications, Department of Engineering Science,\\ 
1-5-1 Chofugaoka, Chofu, Tokyo 182-8585, Japan}

\author{Titus Masese}
\email{titus.masese@aist.go.jp}
\affiliation{Research Institute of Electrochemical Energy (RIECEN), National Institute of Advanced Industrial Science and Technology (AIST), 1-8-31 Midorigaoka, Ikeda, Osaka 563-8577, Japan}
\affiliation{AIST-Kyoto University Chemical Energy Materials Open Innovation Laboratory (ChEM-OIL), Yoshidahonmachi, Sakyo-ku, Kyoto-shi 606-8501, Japan}

\begin{abstract}
We introduce a pseudo-spin model for the argentophilic bond expected in silver-based bilayered materials arising from a spontaneous pseudo-magnetic field interacting with pseudo-spins of two unconventional Ag ions, namely $\rm Ag^{2+}$ and $\rm Ag^{1-}$ electronically distinct from (\textit{albeit} energetically degenerate to) the conventional $\rm Ag^{1+}$ cation typically observed in monolayered materials. This model suggests the possibility of tuning the dimensionality and hence the conductor-semiconductor-insulator properties of honeycomb bilayered materials by application of external fields, analogous to driving a superconducting or Coulomb blockade system to the normal regime by critical magnetic or electric fields respectively.\\

\end{abstract}

\maketitle

\section{Introduction}

\blue{Honeycomb layered frameworks manifesting argentophilic bilayers with the chemical formula ${\rm Ag}_2^{1/2+}M^{3+}\rm O_2^{2-}$ and ${\rm Ag}_3^{2/3+}M_2^{3+}\rm O_4^{2-}$ (where $M$ is a transition metal ions $M^{3+} = \rm Cr^{3+},\, Mn^{3+},\, Fe^{3+},\, Co^{3+},\, Ni^{3+}$ \textit{etc}) have recently become a research subject of great interest.\cite{kanyolo2023honeycomb} Particularly, the perplexing lack of complete geometric frustration of magnetic order on the $M^{3+}$ hexagonal lattice exemplified by a N\'{e}el antiferromagnetic phase transition\cite{yoshida2006spin}, amongst other peculiarities such as magnetoresistance\cite{taniguchi2020butterfly}, has lead some researchers to consider additional mechanisms such as indirect and/or super-exchange Heisenberg interactions, spin canting and/or Dzyaloshinskii-Moriya (anti-symmetric exchange) interactions\cite{yoshida2006spin, sugiyama2006Incommensurate, yoshida2020partially, yoshida2008unique, nozaki2013magnetic, nozaki2008neutron, eguchi2010resonant} (\textit{e.g.} arising from a Jahn-Teller distortion of the octahedral $M^{3+}\rm O_2^{2-}$ structure\cite{yoshida2006spin}) and chirality (\textit{e.g.} arising from the XY model or 120$^\circ$ model\cite{korshunov1986phase}), interpreted to supersede geometric frustration in favour of the N\'{e}el antiferromagnetism.\cite{kanyolo2023honeycomb} 

Meanwhile, the interplay between the highly conductive Ag bilayers and the N\'{e}el antiferromagnetic order has been linked to heavy fermion behaviour (renormalisation of the electron mass leading to an uncharacteristic resistivity versus temperature dependence at low temperatures).\cite{eguchi2010resonant, ji2010orbital, kanyolo2023honeycomb} Moreover, similar to the subflouride $\rm Ag_2^{1/2+}F^{1-}$\cite{tong2011, wang1991, ikezawa1985, fujita1974, andres1966superconductivity, ido1988, ott1928, kawamura1974, argay1966redetermination}, the conduction band structure of the argentophilic bilayers has been typically considered quarter-filled with highly itinerant electrons. Nonetheless, the explanation for the existence of silver subvalency has remained contentious\cite{yin2021reply, lobato2021comment, kovalevskiy2020uncommon, schreyer2002synthesis, johannes2007formation} presumably due to the fact that, until recently\cite{masese2022honeycomb}, its origin was pegged to specific environmental factors within the bilayered framework such as the oxidation state of $M^{3+}$ ions in the honeycomb layered frameworks ${\rm Ag}_2^{1/2+}M^{3+}\rm O_2^{2-}$ and ${\rm Ag}_3^{2/3+}M_2^{3+}\rm O_4^{2-}$\cite{schreyer2002synthesis, johannes2007formation} \red{and other} 
\red{approaches related} 
to polyhedral skeletal electron pair theory (Wade-Mingos molecular clusters\cite{wade1976, mingos1972general, mingos1984}) where the presence of extra localised electrons within the $\rm Ag$ tetrahedral structures of \textit{e.g.} $\rm Ag_{16}^{1/2+}B_4^{3+}O_{10}^{2-}$\cite{kovalevskiy2020uncommon} accounting for the subvalency of $\rm Ag$ is rationalised via \textit{ab initio} density functional theory (DFT) calculations and subsequent electron localisation function analysis.\cite{yin2021reply, lobato2021comment, kovalevskiy2020uncommon} 
 
Recently, a novel class of honeycomb bilayered tellurates ${\rm Ag_6^{1/2+}}M_2\rm TeO_6$ ($M = \rm Ni, Co, Mg, Cu$ \textit{etc} was reported exhibiting the Ag subvalent state, ${\rm Ag}^{v+}$ (experimentally: $1/2+ \leq v \leq 2/3+$; theoretically: $v = 1/2+$)).\cite{masese2022honeycomb} Crucially, a universal framework for \red{the} Ag subvalency was also proposed therein, which relies on $sd$ hybridisation of the $4d_{z^2}$ and $5s$ orbitals of silver resulting in $SU(2)\times U(1)$ gauge interactions between 3 otherwise degenerate valency states of Ag, achieving the $\rm Ag^{1/2+}$ state on a bifurcated honeycomb lattice.\cite{masese2022honeycomb, kanyolo2023honeycomb, kanyolo2022advances2} Particularly, d}\blue{ue to energy proximity of silver $4d^{10}$ and $5s^1$ electrons and environmental factors in silver-based honeycomb bilayered materials such as crystal field splitting of the $4d$ orbitals encouraging $sd$ hybridisation}, the electronic configuration of silver atom ($\rm Ag$) can be in either one of two degenerate states, namely the expected \blue{[Kr]}$4d^{10}5s^1$ configuration \blue{responsible for the} oxidation state $\rm Ag^{1+}$ or the proposed \blue{[Kr]}$4d^95s^2$ configuration.\cite{masese2022honeycomb} \blue{This suggests} two \blue{other} oxidation states exist, namely $\rm Ag^{2+}$ and $\rm Ag^{1-}$, obtained by designating the electrons in either the $5s^2$ or $4d^9$ orbitals \blue{respectively} as the valence electrons \blue{(or \red{potentially} both, in the case of $\rm Ag^{3+}$)}. Whilst these degenerate states \red{are known} 
to exist independently in various materials such as the anti-ferromagnetic material $\rm Ag^{2+}F_2^{1-}$ and \red{in} silver clusters where ${\rm Ag}^{1-}_N$ ($N = 1$)\cite{kurzydlowski2021fluorides, grzelak2017metal, schneider2005unusual, dixon1996photoelectron, ho1990photoelectron, minamikawa2022electron, *minamikawa2022correction}, by a theorem analogous to Peierls\red{'}\cite{lee2011band, peierls1979surprises, garcia1992dimerization}, their co-existence requires \red{the} 
energy degeneracy to be spontaneously broken for the material to ultimately achieve stability.\cite{kanyolo2022advances, kanyolo2022advances2, masese2022honeycomb} 

\begin{figure}
\begin{center}
\includegraphics[width=1.0\columnwidth,clip=true]{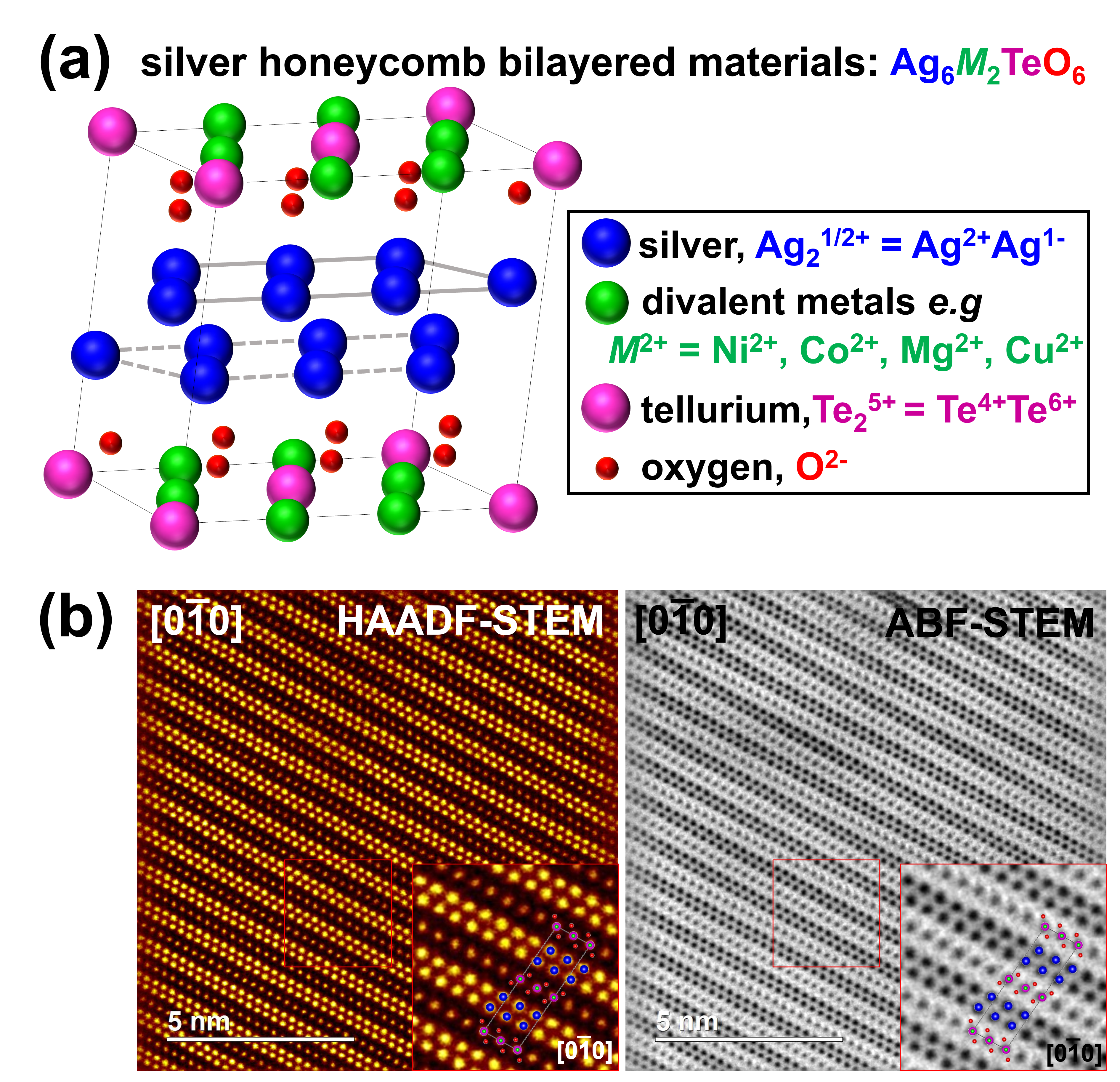}
\caption{Crystalline structure of the recently discovered silver-based honeycomb bilayered material ${\rm Ag_6^{1/2+}}M_2 \rm TeO_6$ where in (a) the valence $1/2+$ silver cations ($\rm Ag_2^{1/2+} = \rm Ag^{2+}Ag^{1-}$) are shown in blue, the divalent metal cations ($M^{2+} = \rm Ni^{2+}, Co^{2+}, Mg^{2+}, Cu^{2+}$ \textit{etc}) are shown in green, tellurium cations $\rm Te_2^{5+} = Te^{4+}Te^{6+}$ are shown in magenta and oxygen anions ($\rm O^{2-}$) are shown in red. (b) Crystalline structure of $\rm Ag_6Ni_2TeO_6$ obtained from the $[0\overline{1}0]$ direction by the high-angle annular dark-field (HAADF) and annular bright-field (ABF) scanning transmission electron microscope (STEM) techniques clearly showing the silver bilayers (inset).\cite{masese2022honeycomb}}
\label{Fig_1}
\end{center}
\end{figure}

For exemplar silver-based bilayered materials not limited to ${\rm Ag_6^{1/2+}}M_2\rm TeO_6$ ($M = \rm Ni, Co, Mg,Cu$ \textit{etc}) whose structure is shown in Figure \ref{Fig_1}, it has been proposed that the coexistence of the silver degenerate states naturally results in anomalous valency states (subvalent states) such as $\rm Ag_2^{1/2+} = Ag^{2+}Ag^{1-}$ or $\rm Ag_3^{2/3+} = Ag^{2+}Ag^{1-}Ag^{1+}$, which also requires the dimerisation of the two silver atoms in each primitive cell of the bipartite honeycomb lattice, enabled by 
\red{an argentophilic bond}.\cite{masese2022honeycomb} It is this dimerisation \red{via argentophilicity} that we propose to be the manifestation of \blue{pseudo-spin interactions with pseudo-magnetic fields} in silver-based bilayered materials. In silver metal, \blue{pseudo-spin} as the explanation for argentophilicity is superfluous since the concept of metallic bonds already suffices. Particularly, metallic bonds arise whenever the valence electrons are extremely delocalised, leading to their collective sharing amongst the metallic ions. Meanwhile, the argentophilic bonds in silver bilayers are not only shorter than silver metallic bonds ($\leq 2.83$\,\,\AA\,\,and $\leq 2.89$\,\,\AA\,\,in bilayered silver structures and elemental silver respectively)\cite{masese2022honeycomb}, but also require the valence electrons to be localised. In the scheme involving silver degenerate states\cite{masese2022honeycomb}, localisation of silver valence electrons is already implied by the involvement of the electronic configuration \blue{[Kr]}$4d^95s^2$ (particularly, $4d_{z^2}^15s^2$) \blue{obtained by} $sd$ hybridisation instead of the conventional $4d^{10}5s^1$ electronic configuration with the usually itinerant $5s^1$ valence electrons localised by the closed-shell pairing in $5s^2$ state. In addition, like in graphene, the pseudo-spins on the bipartite honeycomb lattice are inherited from the behaviour of actual spins of the $4d_{z^2}^1$ orbital electrons\cite{kanyolo2022advances2, georgi2017tuning, mecklenburg2011spin, allen2010honeycomb, kvashnin2014phase} \textit{albeit} the pseudo-spin degree of freedom in silver is expected to clearly manifest for an isolated $4d_{z^2}^1$ orbital due to $4d^9$ crystal field splitting in linear or prismatic coordination to \blue{anions such as} $\rm F^{1-}$ or $\rm O^{2-}$.\cite{masese2022honeycomb}  

\blue{In this paper, we introduce a pseudo-spin model for the argentophilic bond expected in silver-based bilayered materials, arising from a spontaneous pseudo-magnetic field interacting with pseudo-spins of the two Ag degenerate valency states, namely $\rm Ag^{2+}$ and $\rm Ag^{1-}$. The pseudo-magnetic field
\red{is theoretically treated as} a Dirac mass on the honeycomb lattice, interpreted as a bifurcation of the bipartite honeycomb lattice 
\red{resulting in the argentophilic bond}.\cite{masese2022honeycomb, kanyolo2022advances2, kanyolo2023honeycomb} Parity considerations of the fermionic wave function suggest that \red{a} finite magnetic field 
undoes the bifurcation, making pseudo-spin degrees of freedom dual to actual spin degrees of freedom. This implies the possibility of tuning the dimensionality and hence the conductor-semiconductor-insulator properties of silver-based bilayered materials by application of external 
fields, analogous to driving a superconducting or Coulomb blockade system to the normal regime by critical 
magnetic \red{or electric fields respectively.\cite{tinkham2004introduction, kanyolo2021renormalization}} Since the spontaneously generated electron mass likely differs from the actual electron mass, this `renormalisation' mechanism potentially explains the heavy fermion behaviour reported in several silver-based honeycomb layered materials.\cite{eguchi2010resonant, ji2010orbital, kanyolo2023honeycomb} \red{Moreover, m}anifesting such a monolayer-bilayer phase transition experimentally would offer a potent tool to characterise and/or isolate factors related to the $M_2\rm TeO_6$ inter-layer interactions that potentially hinder the realisation of Kitaev physics in materials such as ${\rm Ag}_6M_2\rm TeO_6$ ($M = \rm Co, Ni$, {\textit etc}).\cite{komori2023antiferromagnetic} Finally, the value of this work also lies in the novel mathematical physics tools and ideas (related to emergent gravity in honeycomb layered materials and the \red{Riemann} zeta function\cite{kanyolo2022advances2, kanyolo2020idealised, kanyolo2022cationic}) employed in order to achieve the dual treatment of magnetic and pseudo-magnetic degrees of freedom on the honeycomb lattice.}

Herein, it is prudent to employ units where reduced Planck's constant and speed of mass-less particles in the material are set to unity ($\hbar = \overline{c} = 1$).

\section{Theory}

\begin{figure}
\begin{center}
\includegraphics[width=1.0\columnwidth,clip=true]{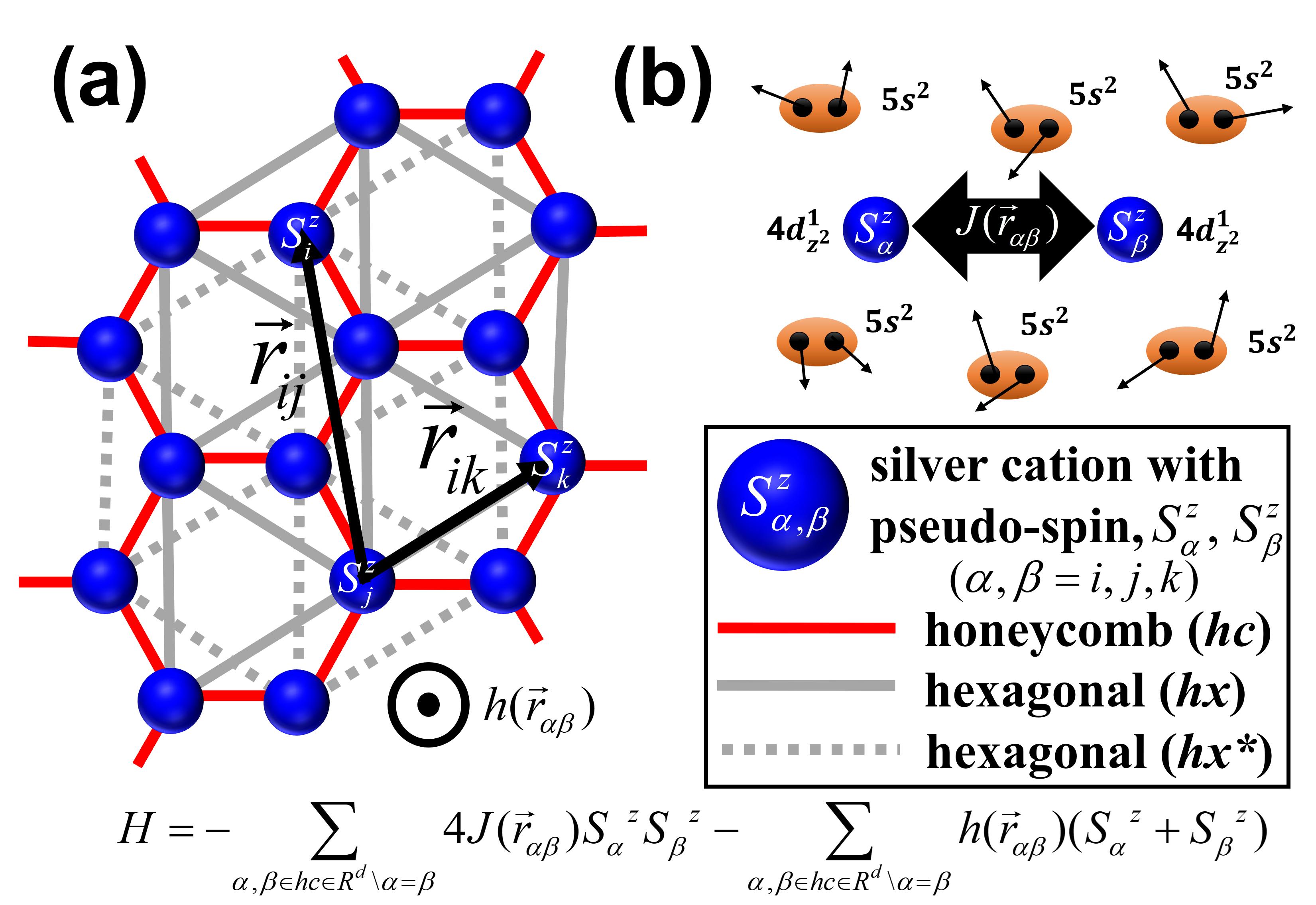}
\caption{Pseudo-spin model introduced on the bipartite $\rm Ag_2^{1/2+} = Ag^{2+}Ag^{1-}$ honeycomb lattice ($hc$) consisting of a pair of equivalent hexagonal lattices ($hx$ and $hx^*$). (a) Vectors $\vec{r}_{ij} \in R^d$ ($d = 2, 3$) and $\vec{r}_{ik} \in R^2$ depicted on the honeycomb lattice spanning between the centres of Ag cations $i$ and $j$ (blue) with pseudo-spins $S_i^z$ and $S_j^z$ respectively, and Ag cations $i$ and $k$ with pseudo-spins $S_i^z$ and $S_k^z$ respectively. The pseudo-spins can interact with a pseudo-magnetic field $h(\vec{r}_{\alpha\beta})$ ($\alpha, \beta = i, j, k$) along the $z$ direction. (b) A depiction of the exchange coupling, $J(\vec{r}_{\alpha\beta})$ between two $4d_{z^2}^1$ magnetic impurities introduced by $\rm Ag^{2+}$ cations at $\alpha$ and $\beta$ with pseudo-spins $S_{\alpha}^z$ and $S_{\beta}^z$, mediated by a sea of itinerant $5s^2$ electrons of $\rm Ag^{1-} = Ag^{1+} + 2e^-$. The total interaction, $H$ excluding terms with $\alpha = \beta$ corresponds to eq. (\ref{Heisenberg_eq1}).}
\label{Fig_2}
\end{center}
\end{figure}

\begin{figure}[!b]
\begin{center}
\includegraphics[width=1.0\columnwidth,clip=true]{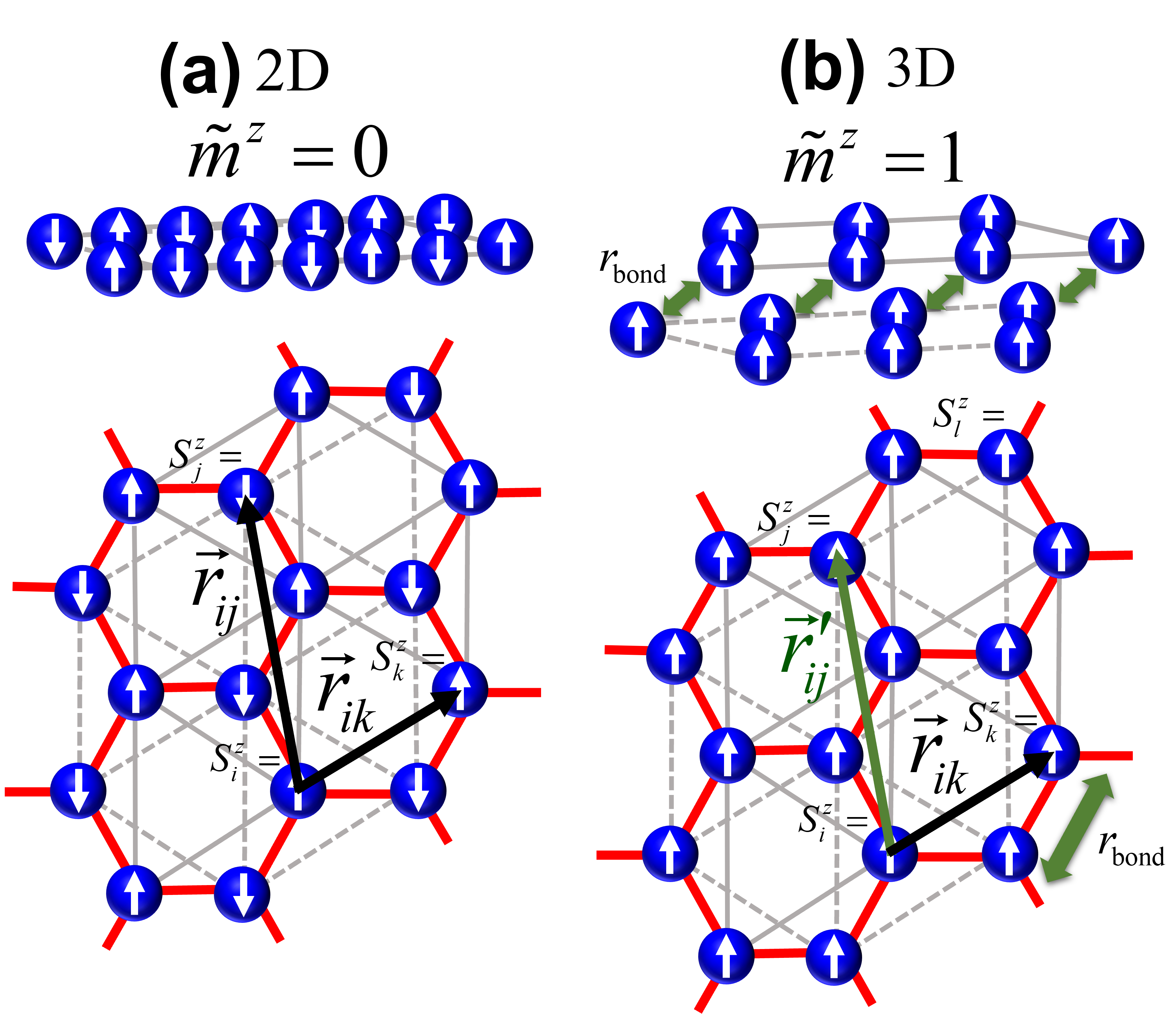}
\caption{Bipartite honeycomb lattice (2D) of silver cations (blue \blue{spheres}) with pseudo-magnetisation $\tilde{m}^z = 0$ and its bifurcated counterpart (3D) with $\tilde{m}^z = 1$, where (a) and (b) depict the vectors and pseudo-spins in Figure \ref{Fig_2}. The vectors in black lack a $z$ component ($\vec{r}_{ij}, \vec{r}_{ik} \in R^2$) whereas the $z$ component of the vector in green is finite ($\vec{r}_{ij} \in R^3$) due to bifurcation\blue{, with $r_{\rm bond}$ and the green double arrow indicating the argentophilic bond}.}
\label{Fig_3}
\end{center}
\end{figure}

\textbf{Pseudo-spin model.}\textemdash \blue{C}onsider $sd$ hybridisation as the interaction between localised spin impurities \blue{$S_{\alpha}^z$ and $S_{\beta}^z$ ($\alpha, \beta = i,j,k$) on a honeycomb lattice} comprising isolated $4d_z^1$ orbitals in the Ag configuration, \blue{[Kr]}$4d^95s^2$, mediated by conduction electrons comprising the $5s^2$ orbitals as illustrated in Figure \ref{Fig_2}. \blue{This leads} to the exchange interaction terms $J(\vec{r}_{ij})$ \blue{between $i$ and $j$ atoms and possible linear terms $h(\vec{r}_{ij})$} in the Heisenberg Hamiltonian\cite{garcia1992dimerization, said1984nonempirical},
\begin{subequations}\label{Heisenberg_eq1}
\begin{align}
    H = \sum_{i, j \in hc \in R^d \setminus i = j}\left [f(\vec{r}_{ij}) + 4J(\vec{r}_{ij})\left (\left\langle S_i^zS_j^z \right\rangle + 1/4 \right )\right ] 
    = \sum_{i, j \in hc \in R^d \setminus i = j} \left [4J(\vec{r}_{ij})\left\langle S_i^zS_j^z \right\rangle + h(\vec{r}_{ij})\left \langle (S^z_i + S^z_j) \right \rangle\right ],
\end{align}
under the constraint, 
\begin{align}
    f(\vec{r}_{ij}) = - J(\vec{r}_{ij}) + h(\vec{r}_{ij})\left \langle (S^z_i + S^z_j) \right \rangle,
\end{align}
\end{subequations}
where the sum is evaluated over silver atoms $i$ and $j$ excluding $i \neq j$ in a potentially bifurcated bipartite honeycomb hexagonal lattice ($hc$) comprising a pair of hexagonal lattices ($hx, hx^*$) displaced along the $z$ direction in Euclidean space $R^d$  \blue{(bifurcated honeycomb lattice)} when $d = 3$ \blue{dimensions (3D)}, and non-bifurcated for $d = 2$ \blue{dimensions (2D)} in the plane, $R^2$, as displayed in Figure \ref{Fig_3}. Here, $f(\vec{r}_{ij})$, $J(\vec{r}_{ij})$ and $h(\vec{r}_{ij})$ are yet unknown energy functions of the relative positions $\vec{r}_{ij} = \vec{r}_i - \vec{r}_j$ of next neighbour Ag atoms $i, j$ at positions $\vec{r}_i$ and $\vec{r}_j$ respectively, and $S^z_i$ or $S^z_j$ are the $z$ direction spin operators for pseudo-spin impurities at $i, j$ ($4d_{z^2}^1$ electron spins in each $4d^95s^2$ $\rm Ag$ atom) within a sea of conduction electrons. In this picture, the bond length is given by the silver separation distance, $r_{ij} = |\vec{r}_{ij}| \equiv r_{\rm bond}$ when $H = 0$. Thus, figuring out the functional form of the unknown energy functions on the honeycomb lattice which satisfy $H = 0$ either \blue{analytically} or via \textit{ab initio} calculations amounts to solving the bifurcation problem\blue{, whose results can then be compared with experiment}. In this endeavour, next-neighbour summation approximation may be taken provided it is justified on physical grounds. While particularly unwieldy, similar computational challenges have been overcome for the analogous problem in $d = 1$ dimensions (1D), corresponding to the case of dimerisation of carbon atoms in polyacetylene and other similar conjugated hydrocarbons.\cite{garcia1992dimerization, said1984nonempirical} 

Nonetheless, we are solely interested in pursuing the analytic behaviour of eq. (\ref{Heisenberg_eq1}), starting with the simplest case of the two-body problem. The next-neighbour pseudo-magnetisation and pseudo-spin correlation expectation values are given by, 
\begin{subequations}\label{pseudo_eq}
\begin{align}
    \tilde{m}^z = \left \langle (S^z_i + S^z_j)\right \rangle = 0, \pm 1,\\
    \tilde{c}^z = \left \langle S_i^zS_j^z \right \rangle = -1/4, +3/4,
\end{align}
respectively, which can be associated with the pseudo-spin wave functions as follows,
\begin{align}\label{pseudo_H_eq}
\begin{matrix}
\underline{{\rm wave\,\,function}, \blue{\Sigma_{ij}}} & \underline{\rm magnetisation} & \underline{\rm correlation} \\
\frac{1}{\sqrt{2}}\left (|\uparrow\,\downarrow\,\rangle - |\downarrow\,\uparrow\,\rangle \right ), & \tilde{m}^z = 0, & \tilde{c}^z = -1/4, \\ 
\frac{1}{\sqrt{2}}\left (|\uparrow\,\downarrow\,\rangle + |\downarrow\,\uparrow\,\rangle \right ), & \tilde{m}^z = 0, & \tilde{c}^z = + 3/4,\\
|\uparrow\,\uparrow\, \rangle, & \tilde{m}^z = 1, & \tilde{c}^z = + 3/4,\\ 
|\downarrow\,\downarrow\, \rangle, & \tilde{m}^z = -1, & \tilde{c}^z = + 3/4.
\end{matrix}    
\end{align}
\end{subequations}
It is prudent to also consider the actual spin magnetisation and correlations in the $z$ direction given by,
\begin{subequations}\label{actual_eq}
\begin{align}
    m^z = \left \langle (s^z_i + s^z_j)\right \rangle = 0, \pm 1,\\
    c^z = \left \langle s_i^zs_j^z \right \rangle = -1/4, +3/4,
\end{align}
respectively where $s_i, s_j$ are the actual spin operators, which can be associated with the actual spin wave functions as follows,
\begin{align}\label{actual_H_eq}
\begin{matrix}
\underline{{\rm wave\,\,function}, \blue{\sigma_{ij}}} & \underline{\rm magnetisation} & \underline{\rm correlation} \\
\frac{1}{\sqrt{2}}\left (|u\,d\,\rangle - |d\,u\,\rangle \right ), & m^z = 0, & c^z = -1/4, \\ 
\frac{1}{\sqrt{2}}\left (|u\,d\,\rangle + |d\,u\,\rangle \right ), & m^z = 0, & c^z = + 3/4,\\
|u\,u\, \rangle, & m^z = 1, & c^z = + 3/4,\\ 
|d\,d\, \rangle, & m^z = -1, & c^z = + 3/4.
\end{matrix}    
\end{align}
\end{subequations}
\blue{Here, $|u \rangle$ and $|d \rangle$ are respectively the spin up and down wave functions of itinerant $5s^2$ silver conduction electrons, to be distinguished from the pseudo-spin up and down wave functions, $|\uparrow \, \rangle$ and $|\downarrow \, \rangle$ respectively considered to originate from the spins of localised $4d_{z^2}^1$ silver electrons.} Thus, using the notation, $\Sigma_{ij} = (\tilde{m}^z, \tilde{c}^z)_{ij}$ to label the pseudo-spin wave functions and $\sigma_{ij} = (m^z, c^z)_{ij}$ to label the actual spin wave functions, we can consider the parity of the spatial wave functions in order to extract valuable information about the ground state of the pseudo-spin system given by eq. (\ref{Heisenberg_eq1}). 

In particular, the spatial \blue{two-particle} wave function $\psi_{\pm}(\vec{r}_{ij})$ of \blue{adjacent silver} fermions in the absence of pseudo-spin degrees of freedom must satisfy,
\begin{align}\label{wave function_eq}
    \psi_{\pm}(\vec{r}_{ij})\sigma^{\mp}_{ij} = -\psi_{\pm}(\vec{r}_{ji})\sigma^{\mp}_{ji} 
    = \frac{1}{2}\left [\phi_i(\vec{r}_i)\phi_j(\vec{r}_j) \pm \phi_i(\vec{r}_j)\phi_j(\vec{r}_i)\right ]\sigma^{\mp}_{ij},
\end{align}
implying $\sigma^{\mp}_{ij} = \mp \sigma^{\mp}_{ji}$, where $\phi_i(\vec{r}), \phi_j(\vec{r})$ is the spatial wave function of a conduction electron at $\vec{r}_i, \vec{r}_j$ hopping between Ag cations on adjacent lattice sites and $\vec{r}_{ij} = \vec{r}_i - \vec{r}_j$. However, assuming the honeycomb lattice introduces the pseudo-spin degree of freedom with wave functions, $\Sigma_{ij}$ the composite \blue{two-silver} wave function can be written in a generic form as, 
\begin{align}\label{ground_state_eq}
    \psi^p(\vec{r}_{ij})\Sigma_{ij}^a\sigma_{ij}^b = -\psi^p(\vec{r}_{ji})\Sigma_{ji}^a\sigma_{ji}^b 
    = \frac{1}{2}\left [\phi_i(\vec{r}_i)\phi_j(\vec{r}_j) + p\phi_i(\vec{r}_j)\phi_j(\vec{r}_i)\right ]\Sigma_{ij}^a\sigma_{ij}^b,
\end{align}
where $p = \pm$, $a = \pm$ and $b = \pm$ indicate the symmetry ($+$)/anti-symmetry ($-$) of the respective wave functions. \blue{For instance, for plane wave solutions, the symmetry/anti-symmetry of the spatial wave function corresponds to even/odd parity. This is evident for a model comprising left- and right-moving plane waves of conduction electrons hopping between $\rm Ag^{2+}$ and $\rm Ag^{1-}$ along a honeycomb edge,
\begin{align}
   \blue{
   \phi_{\alpha}(\vec{r}_{\alpha}) = \frac{1}{\sqrt{\Omega}}\exp(\#\sqrt{-1}\, \vec{k}_{\rm F}\cdot\vec{r}_{\alpha})
   },
\end{align}
where $\vec{k}_{\rm F}$ is the Fermi wave vector, \blue{$\alpha = i, j$} and their chirality is given by $\# = \pm$, where $\Omega$ is the wave function normalisation factor. Requiring next-neighbour interacting $\rm Ag$ states to have opposite chirality inherited from the conduction electrons (a requirement for the generation of a mass term between $\rm Ag^{2+}$ and $\rm Ag^{1-}$\cite{masese2022honeycomb}), we obtain plane wave solutions of the spatial wave function,
\begin{align}\label{wave function_eq2}
\phi^p(\vec{r}_{ij}) = 
   \begin{cases}
    \,\,\,\,\Omega^{-1}\cos(\vec{k}_{\rm F}\cdot\vec{r}_{ij}), \,\,\,\,p = +,\\
     \,\,\,\,\Omega^{-1}\sqrt{-1}\sin(\vec{k}_{\rm F}\cdot\vec{r}_{ij}), \,\,\,\,p = -,
\end{cases}
\end{align}
appearing in eq. (\ref{wave function_eq}).}

Proceeding, the only possible signs for a pair of \blue{silver} fermions are given by the following cases: 
\begin{align}\label{cases_eq}
\begin{matrix}
\underline{\rm cases}: & p & a & b \\
\blue{\rm (i):} & + & + & - \\
\blue{\rm (ii):} & + & - & + \\
\blue{\rm (iii):} & - & - & - \\
\blue{\rm (iv):} & - & + & + .
\end{matrix} 
\end{align}
Moreover, \blue{assuming potentials $h(\vec{r}_{ij})$ and $J(\vec{r}_{ij})$ are symmetric in $\vec{r}_{ij}$ and the Hamiltonian $H$ commutes with the parity operator,} the odd parity states ($p = -$) in eq. (\ref{cases_eq}) are forbidden by \blue{a known theorem that the ground state is} even parity\blue{.\cite{sakurai1995modern} Thus,} we should cross out all the $p = -$ states\blue{, which} leaves the only viable \blue{ground states} \blue{in eq. (\ref{cases_eq}) as cases (i) and (ii)}. Note that, for vanishing respective magnetic fields, each of these ground states are 3-fold degenerate. However, degeneracy of the ground state is forbidden by a theorem by Peierls, which guarantees that the lattice must distort to lift the degeneracy.\cite{lee2011band, peierls1979surprises, garcia1992dimerization} \blue{Identifying} the distortion on the honeycomb lattice \blue{with the} finite pseudo-magnetic interaction, $h(\vec{r}_{ij}) \neq 0$, we find \blue{case (i) as} the unique ground state, where the pseudo-magnetic interaction (\blue{responsible for the} bifurcation of the monolayered silver honeycomb lattice into bilayers\cite{masese2022honeycomb}) must appear spontaneously thus breaking the degeneracy. On the other hand, a finite actual magnetic field selects instead the (other) ground state \blue{given by case (ii)} in eq. (\ref{cases_eq}), which is guaranteed to be non-degenerate (by the presence of the magnetic field). 

Consequently, due to the wave function symmetry considerations \blue{the} ground state \blue{given by case (ii) in eq. (\ref{cases_eq})} must be without strain or distortions ($h(\vec{r}_{ij}) = 0$), implying a finite external magnetic field is present, 
\begin{align}\label{B_eq}
    B \equiv (\vec{\nabla}\times\vec{A})_z \neq 0,
\end{align}
which tunes the argentophilic bond ($\vec{A}$ is the electromagnetic vector potential) and hence the monolayer-bilayer phase transition, thus altering the dimensionality of the Ag lattice from 3D \blue{back} to 2D, corresponding to the displayed lattices in Figure \ref{Fig_3}. Intuitively, the pseudo-magnetic interaction is analogous to the order parameter in the Ginzburg-Landau or the energy gap in Bardeen-Cooper-Schrieffer (BCS) theories of superconductivity, which are related and can be tuned by an external magnetic field driving the superconducting state to normal.\cite{tinkham2004introduction}\\ 

\textbf{Paired electrons.}\textemdash \blue{T}he time component of the \blue{conduction electron} wave functions $\phi_{i,j}(\vec{r}_{i,j})$ \blue{or the overall silver ground state wave function} is the phase \blue{factors}, $\exp(-iE_{\pm}t)$ where the dispersion relation at any honeycomb vertex with a $\rm Ag^{2+}$ or $\rm Ag^{1-}$ \blue{ion} satisfies the linear dispersion relation, $E_{\pm} = \pm |\vec{k}_{\rm F}|$ \blue{and} $\pm$ reflects the electron($+$)/hole($-$) energy dispersion relations. Thus, the Green's function of interest is the appropriately defined Fourier transform of the phase factor\cite{mattuck1992guide}, 
\begin{align}
    \mathcal{G}_{\pm}(\omega) = \frac{1}{\omega - E_{\pm} + i\epsilon}
    = -i\int_{-\infty}^{\infty}dt\,\exp(i\omega t - \epsilon t)\theta(t)\exp(-iE_{\pm}t),
\end{align}
where $i = \sqrt{-1}$, $\theta(t)$ is the Heaviside step function restricting the Fourier transform to the range $0 \leq t \leq \infty$ and $\epsilon \simeq 0$ is the infinitesimal with the properties $\epsilon \times \infty = \infty$ and $\epsilon\times 0  = 0$. In subsequent calculations \blue{using the Green's function}, we shall set $\epsilon = 0$. 

Now, departing from \blue{only considering adjacent Ag ions in} the two-body problem, \blue{but} assuming that the \blue{adjacent} pseudo-spins on the honeycomb lattice are \textit{anti-parallel} as shown in Figure \ref{Fig_3}(a), the modification of the electron dispersion relation by interactions with the pseudo-magnetisation component of $H$ in eq. (\ref{Heisenberg_eq1}) \blue{identically} vanishes since, 
\begin{multline}\label{R2_eq}
    U = \sum_{i, j \in hc \in R^2 \setminus i = j} h(\vec{r}_{ij})\left \langle (S^z_i + S^z_j) \right \rangle\\
    = \sum_{i, j \in hx \setminus i = j} h(\vec{r}_{ij})\left \langle (S^z_i + S^z_j) \right \rangle + \sum_{i, j \in hx^* \setminus i = j} h(\vec{r}_{ij})\left \langle (S^z_i + S^z_j) \right \rangle 
    + \sum_{i, j \in hc \in R^2 \setminus i = j, \vec{r}_{ij} \in hx, hx^*} h(\vec{r}_{ij})\left \langle (S^z_i + S^z_j) \right \rangle\\
    = \sum_{i, j \in hx \setminus i = j} h(\vec{r}_{ij})\times 1 + \sum_{i, j \in hx^* \setminus i = j} h(\vec{r}_{ij})\times -1
    + \sum_{i, j \in hc \in R^2 \setminus i = j, \vec{r}_{ij} \in hx, hx^*} h(\vec{r}_{ij})\times 0 = 0.
\end{multline}
On the other hand, considering \blue{adjacent} pseudo-spins to be \textit{parallel} instead (\textit{e.g.} pseudo-spin \textit{up} commensurate with the bifurcated lattice shown in Figure \ref{Fig_3}(b)), only the vectors not within the 2D hexagonal lattices $\vec{r}_{ij} \notin R^2$ (\textit{i.e.} $i,j \in hc \in R^3 \setminus i \neq j, \vec{r}_{ij} \in R^2$) will have a finite $z$ component,  
\begin{multline}\label{R3_eq}
     U = \sum_{i, j \in\,\,hc\,\,\in R^3 \setminus i = j} h(\vec{r}_{ij})\left \langle (S^z_i + S^z_j) \right \rangle 
     = \sum_{i, j \in hx, hx^*\in R^2 \setminus i = j} h(\vec{r}_{ij})\times 1 
     + \sum_{i,j \in hc \in R^3 \setminus i \neq j, \vec{r}_{ij} \in R^2} h(\vec{r}_{ij})\times 1\\
     \equiv U_{R^2} + U_{R^3}. 
\end{multline}
Due to the finite $z$ component, the \blue{second} sum of terms \blue{given by $U_{R^3}$} in eq. (\ref{R3_eq}) together with the sum involving the functions $J(\vec{r}_{ij})$ in eq. (\ref{Heisenberg_eq1}) lead to the expression given in eq. (\ref{orbits_eq}) for the argentophilic bond. Meanwhile, the \blue{first} sum of terms in eq. (\ref{R3_eq}) \blue{given by $U_{R^2}$} leads to a mass gap for 2D electron \blue{(silver) dynamics} on the bifurcated lattice. 

To see this, we shall assume $|U_{R^2}|$ takes the same functional form in position as well as momentum space, which we \blue{can} justify using eq. (\ref{Theta_eq}). Intuitively, this arises from the fact that the reciprocal of the hexagonal lattice is another hexagonal \textit{albeit} with unit vectors rotated by $\pi/2$ and appropriately re-scaled momentum vectors. Now, we wish to calculate the effective Green's function due to interactions of $\mathcal{G}_+(\omega)$, $\mathcal{G}_-(\omega)$ and $U_{R^2}$ given by,
\begin{subequations}
\begin{multline}
    \mathcal{G}_+^{\rm eff}(\omega) \equiv \mathcal{G}_+ + \mathcal{G}_+U_{R^2}\mathcal{G}_-U_{R^2}^*\mathcal{G}_+ 
    + \mathcal{G}_+U_{R^2}\mathcal{G}_-U_{R^2}^*\mathcal{G}_+U_{R^2}\mathcal{G}_-U_{R^2}^*\mathcal{G}_+ + \cdots\\
    = \frac{1}{\mathcal{G}_+^{-1} - |U_{R^2}|^2\mathcal{G}_-} = \frac{\omega - E_-}{\omega^2 - E_{\pm}^2 - |U_{R^2}|^2},
\end{multline}
where we have used the generic expression 
\red{$w(1 + (wz) + (wz)^2 + \cdots) = 1/(w^{-1} - z)$} under a regularisation scheme for values of 
\red{$z \geq 1$} where $z, w$ can be arbitrary complex functions.\cite{mattuck1992guide} We can thus describe the \blue{electron (silver)} interactions alongside their chiral degrees of freedom as four-spinors by \blue{a} diagonalisation \blue{scheme} into quasi-particle interactions of the form, 
\begin{align}
    \mathcal{G}_+^{\rm eff}(\omega) = \frac{\omega - E_-}{\omega^2 + \mathcal{E}_-\mathcal{E}_+} = \frac{\mu_+^2}{\omega - \mathcal{E}_+} + \frac{\mu_-^2}{\omega - \mathcal{E}_-} 
    = \frac{\mu_+^2}{\omega - \vec{k}_{\rm F}\cdot\vec{\alpha} - |U_{R^2}|\gamma^0} + \frac{\mu_-^2}{\omega + \vec{k}_{\rm F}\cdot\vec{\alpha} + |U_{R^2}|\gamma^0},
\end{align}
\end{subequations}
amounting to the result obtained by a Bogoliubov transformation of the particle/quasi-particle fermionic (creation, annihilation) operators ($c^{\dagger}, c$)/($d^{\dagger}, d$) respectively\cite{mattuck1992guide} with $\mathcal{E}_{\pm} = \pm \sqrt{E_{\pm}^2 + |U_{R^2}|^2} = \pm (\vec{k}_{\rm F}\cdot\vec{\alpha} + |U_{R^2}|\gamma^0)$ and, 
\begin{align}\label{coherence_eq}
    \mu_{\pm} = \sqrt{\frac{1}{2}\left (1 \pm \frac{E_{\pm}}{\mathcal{E}_{\pm}}\right )}, 
\end{align}
are the so-called BCS coherence factors in the Bogoliubov transformation, 
\begin{align}
    d^{\dagger} = \mu_+c^{\dagger} + \mu_- c,\\
    d = \mu_+ c + \mu_- c^{\dagger}.
\end{align}
Note that, $c^2 = (c^{\dagger})^2 = d^2 = (d^{\dagger})^2 = 0$, $d^{\dagger}d + dd^{\dagger} = c^{\dagger}c + cc^{\dagger} = \mu_-^2 + \mu_+^2 = 1$, and $\gamma^0$ and $\vec{\alpha}$ are $4\times 4$ matrices related to Dirac matrices $\gamma^{\mu} = \gamma^0(1 , \vec{\alpha})$ which satisfy the Clifford algebra $\gamma_{\mu}\gamma_{\nu} + \gamma_{\nu}\gamma_{\mu} = 2\eta_{\mu\nu}$ with \blue{$\gamma_{\mu} = \eta_{\mu\nu}\gamma^{\nu}$ and} $\eta_{\mu\nu}$ the Minkowski space-time metric tensor. Finally, it is clear that: 
\begin{enumerate}
    \item The Dirac mass corresponds to $m_{\Delta(d)} = |U_{R^2}|$ which must vanish at the critical point of the phase transition such as in eq. (\ref{R2_eq}) even for $h(\vec{r}_{ij}) \neq 0$;
    \item Moreover, \blue{from the conclusions using the parity argument above,} $U_{R^2}$ ought to depend on external magnetic fields 
    tuning it in eq. (\ref{R3_eq}) leading to $U_{R^2} = 0$;
    \item This tuning is expected to be possible even when the bifurcated lattice has $g - 1$ vacancies extracted by external electric fields, where $g$ is defined as the genus of an emergent manifold within the context of an idealised model\cite{kanyolo2022advances2, kanyolo2020idealised, kanyolo2022cationic};
    \item The expression expected for the mass term is $m_{\Delta(d)} \equiv 2m\Delta(d)$, with $m$ a constant with dimensions of mass/energy and $\Delta(d) = (d - 2)/2$ the conformal dimension for mass-less scalar fields.\cite{masese2022honeycomb, kanyolo2022advances2}
\end{enumerate}
Thus, determining an appropriate function $h(\vec{r}_{ij})$ in eq. (\ref{R3_eq}) manifesting these properties, alongside $J(\vec{r}_{ij})$ is tantamount to \blue{finding} a solution for the functional form of the argentophilic bond in the bifurcated honeycomb lattice.\\

\textbf{Proposed solution.}\textemdash Introducing the lattice constant $a$, the vectors in the hexagonal lattices satisfy $r_{ij}^2/a^2 = 2n$, where $n \in \mathbb{N}$ is a finite positive integer. Moreover, we shall consider a trial function for the pseudo-magnetic field,
\begin{align}\label{pseudo_h_eq}
    h(\vec{r}_{ij}) = -\frac{m}{C}\left (\frac{\sqrt{2}\,a}{r_{ij}} \right )^{2(\hat{s} + \hat{s}^*)},
\end{align}
consistent with an emergent Liouville conformal field theory (CFT)\cite{kanyolo2022advances2}, 
\begin{multline}\label{define_h_Eq}
   -\frac{m}{C}\sum_{i, j \in hx, hx^*\in R^2 \setminus i = j} \exp(2(\hat{s} + \hat{s}^*)\Phi(\vec{r}_{ij})) 
   = -\frac{m}{C}\sum_{i, j \in hx, hx^*\in R^2 \setminus i = j}\frac{(\sqrt{2}\,a)^{2(\hat{s} + \hat{s}^*)}}{r_{ij}^{2(\hat{s} + \hat{s}^*)}} \equiv U_{R^2}(\hat{s}, \hat{s}^*)\\
   = \sum_{i, j \in hx, hx^*\in R^2 \setminus i = j} h(\vec{r}_{ij}) = -\frac{2m}{C_{\rm eff}}\sum_{n = 1}^{\infty}\frac{C_n}{n^{\hat{s} + \hat{s}^*}},
\end{multline}
with $1/C$ a proportionality constant, where $\Phi(\vec{r}_{ij}) = -\frac{1}{2}\ln (r_{ij}^2/(\sqrt{2}\,a)^2)$ is the Liouville field defined on the hexagonal lattices\cite{kanyolo2022advances2}, the potential $U_{R^2}(\hat{s}, \hat{s}^*)$ is comprised solely of the sum over silver atoms $i,j$ connected only by 2D vectors $\vec{r}_{ij} \in R^2$ in the hexagonal lattices \textit{i.e.} $i,j \in hx, hx^* \in R^2\ i = j$ (eq. (\ref{R3_eq})) with the lattice now assumed to extend to infinity in $R^2$, $\hat{s}, \hat{s}^*$ are scaling parameters to be solved for, which shall determine the mass term proportional to $U_{R^2}$ and the coefficient $C_n$ is the number of vectors of norm $2n$ on each hexagonal lattice. \blue{T}he effective proportionality constant $1/C_{\rm eff} \propto 1/C$ arises from translating the sum over the silver atoms $i,j$ to the sum over vectors $\vec{r}_{ij} \in hx, hx^* \in R^2$. 

\begin{figure}
\begin{center}
\includegraphics[width=0.6\columnwidth,clip=true]{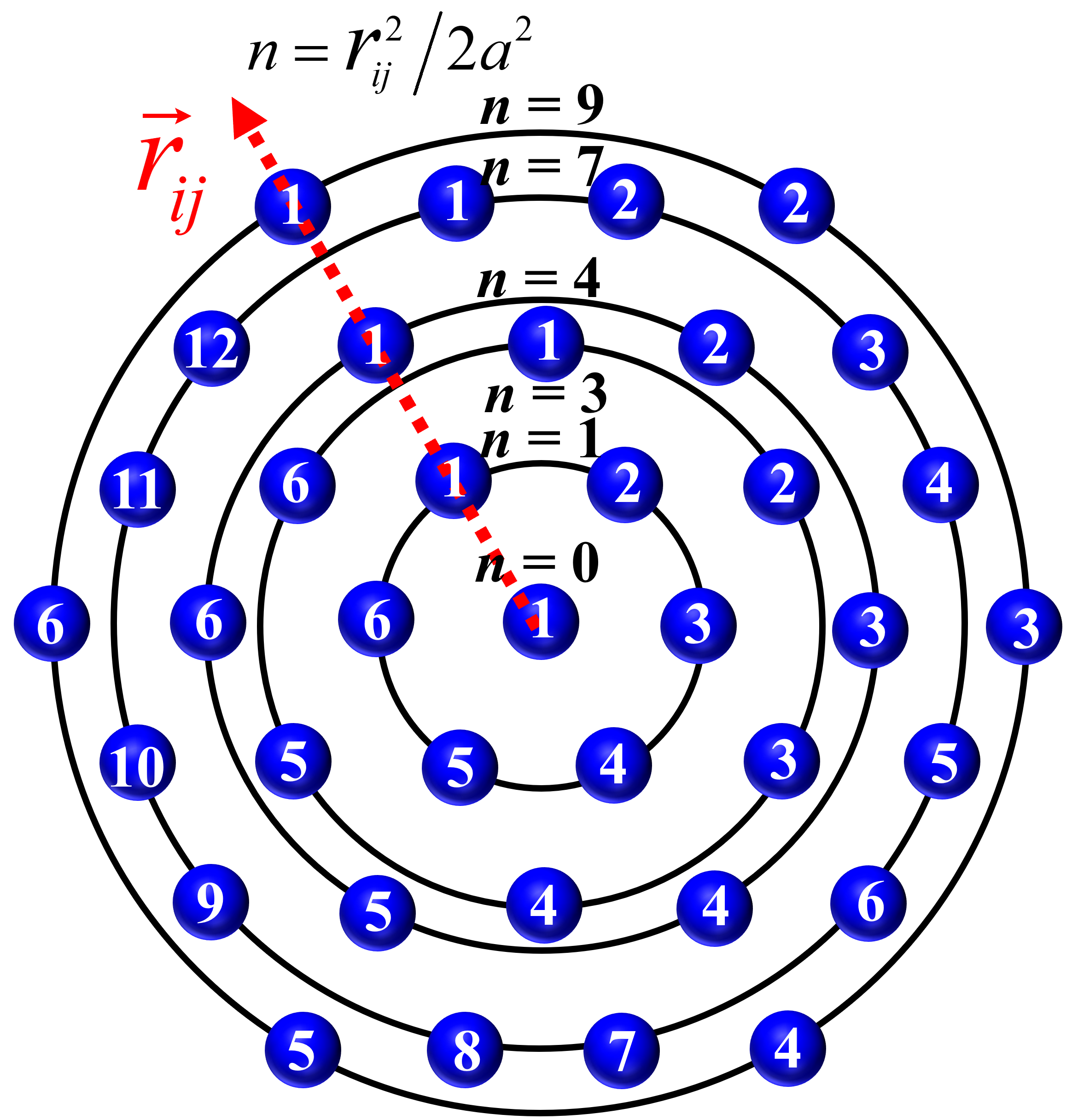}
\caption{Hexagonal lattice of $\rm Ag$ cations (blue) a distance $r_{ij} = a\sqrt{2n}$ radially apart from the origin ($n = 0$), where $a$ is the lattice constant. The concentric black circles correspond to $n = 1, 3, 4, 7, 9 \cdots$ with $n \rightarrow \infty$ assumed in calculations. The numbering in white is availed for ease of counting the number of vectors at a particular radial distance from the origin, displayed in eq. (\ref{C_n_eq}).}
\label{Fig_4}
\end{center}
\end{figure}

Notably, the coefficients $C_n$ can be generated by the theta function of the hexagonal lattice, 
\begin{subequations}\label{Theta_eq}
\begin{align}
    \Theta_{\Lambda}(\tau) = \sum_{\vec{r}_{ij} \in \Lambda} \exp\left (i\pi\tau\frac{r_{ij}^2}{a^2}\right )
    = \sum_{n = 0}^{\infty}C_nq^n(\tau),
\end{align}
where $\Lambda = hx \,\,{\rm or}\,\, hx^*$, $q(\tau) = \exp(2\pi i\tau)$, $\tau \in \mathbb{H}_+$ is complex-valued and restricted to the upper-half plane, $\mathbb{H}_+$ and $r_{ij} = |\vec{r}_{ij}|$ is displayed in Figure \ref{Fig_4}. Thus, we can use the Poisson summation formula\cite{pinsky2008introduction, kanyolo2022advances2} to yield the functional form of $\Theta_{\Lambda'}$ in momentum space ($\vec{k}_{ij} \in \Lambda'$), 
\begin{align}
     \Theta_{\Lambda}(-1/\tau) = -i\tau \Theta_{\Lambda'}(\tau),\\
     \Theta_{\Lambda'}(\tau) = \sum_{\vec{k}_{ij} \in \Lambda'} \exp\left (i\pi\tau\frac{k_{ij}^2}{k_0^2}\right )
    = \sum_{n = 0}^{\infty}C_nq^n(\tau),
\end{align}
\end{subequations}
with $|\vec{k}_{ij}| = k_{ij}$, $k_0$ the lattice constant of $\Lambda'$ satisfying $k_{ij}^2/2k_0^2 = n \in \mathbb{N}$ and $\vec{k}_{ij} \in \Lambda' = {\rm dual}(hx) \,\,{\rm or}\,\, {\rm dual}(hx^*)) = hx' \,\,{\rm or}\,\, hx^{'*}$. \blue{Thus, due to the same functional form between the lattice theta functions $\Theta_{\Lambda}(\tau)$ and $\Theta_{\Lambda'}(\tau)$, their Mellin transform under a change of variable $\tau = i\beta/2\pi$, is proportional to the potential $U_{R^2}$, taking the same functional form in position and momentum spaces.\cite{kanyolo2022advances2}}  

\begin{figure}
\begin{center}
\includegraphics[width=\columnwidth,clip=true]{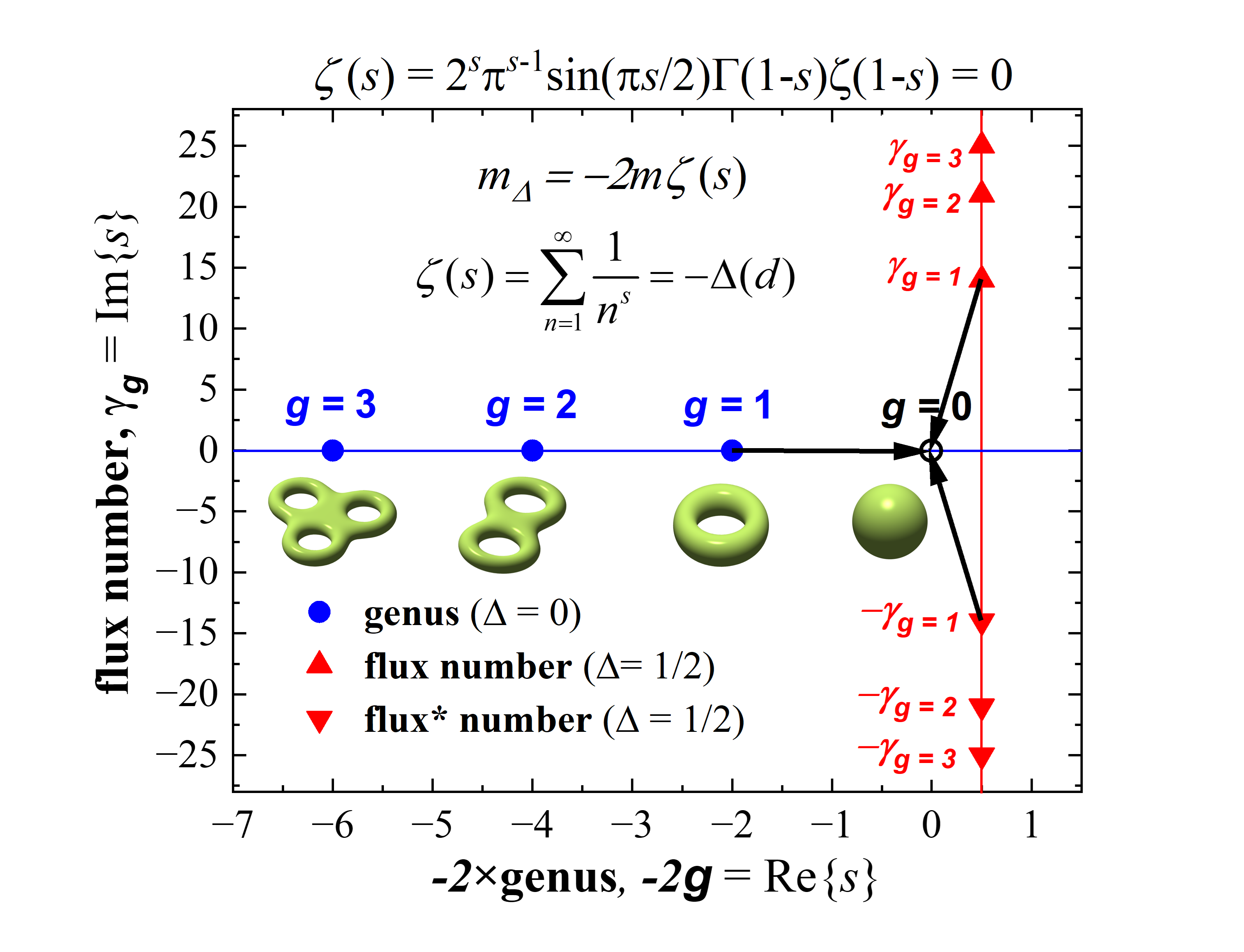}
\caption{Plot of \blue{genus} versus magnetic flux numbers corresponding to a scale invariant $m_{\Delta} = 0$ theory on a bipartite lattice, suggested to be a good approximation for the case of the bifurcated bipartite honeycomb lattice in silver bilayered materials. Since the \blue{generated} masses are proportional to the Riemann zeta function (eq. \ref{continuum_h_Eq2}), this suggests scale invariant theories with $\zeta(s) = 0$, namely Liouville CFT and Chern-Simons theory (eq. (\ref{Idealised_eq})) on Riemannian manifolds $\mathcal{A}_g$ shown in green \blue{with genus $g$}, live at the trivial and essential zeros respectively, as depicted by the blue dots and red triangles. The essential zeros \blue{plotted} were approximated to nearest integers\cite{Sloane1964Nearest}, for ease of plotting. The triple point \blue{of the phase diagram} at $g = 0$ (unshaded black dot) \blue{indicated} by the black arrows corresponds to the condition for stable silver bilayers.}
\label{Fig_5}
\end{center}
\end{figure}

Proceeding, the summation in eq. (\ref{define_h_Eq}) yields expressions where the first few coefficients are given by\cite{Sloane1964Theta},
\begin{align}\label{C_n_eq}
\begin{tabular}{c|ccccccccccc}
    $n$ & 0 & 1 & 2 & 3 & 4 & 5 & 6 & 7 & 8 & 9 & $\cdots$\\
    \hline
    $C_n$ & 1 & 6 & 0 & 6 & 6 & 0 & 0 & 12 & 0& 6& $\cdots$
\end{tabular}\,\,\,\,\,\,\,.
\end{align}
Nonetheless, we are interested in the continuum \blue{approximation}, $C_n \rightarrow C_{n + 1}$ and set $C_{\rm eff} = C_n$ to yield,
\begin{align}\label{continuum_h_Eq2}
   U_{R^2}(\hat{s}, \hat{s}^*) \rightarrow -2m\sum_{n = 1}^{\infty}\frac{1}{n^s} = -2m\zeta(s), 
\end{align}
where $\zeta(s)$ is the Riemann zeta function and $s = \hat{s} + \hat{s}^*$. Famously, $\zeta(s)$ has an analytic continuation,
\begin{align}\label{analytic_eq}
    \zeta(s) = 2^{s}\pi^{s-1}\sin\left( \frac{\pi s}{2} \right)\Gamma(1 - s)\zeta(1 - s), 
\end{align}
in the complex plane $s = \delta + i\gamma$ for real values of $\delta$ and $\gamma$, especially relevant in finding \blue{its} values for $\delta < 1$ with,
\begin{align}\label{Gamma_F_eq}
    \Gamma(s) = \int_0^{\infty} \frac{d\beta}{\beta}\beta^{s}\exp(-\beta),
\end{align}
the Gamma function. \blue{Thus}, we find, 
\begin{align}\label{zeta_eq}
-\zeta(s) = \Delta(d) = 
\begin{cases}
\,\,\,\,\,\,0 \,\,\,\,\,\,\,\,\, (s = -2g \neq 0)\\
\,\,\, 1/2 \,\,\,\,\,\,\,\,\,\,\,\,\,\,\,\,\,(s = 0)
\end{cases},
\end{align}
where $\Delta(d) = (d - 2)/2$, $g$ is taken to be the genus of an emergent 2D manifold $\mathcal{A}_g$ in the Liouville CFT and $-2g \neq 0$ are the so-called trivial zeros of $\zeta(s)$, as shown in Figure \ref{Fig_5}. 

Transitions between these $g \in \mathbb{N}$ states labelling the number of vacancies $\nu = g - 1 \geq 0$ or equivalently the genus of the emerging manifold $\mathcal{A}_g$ with Gaussian curvature $K$ satisfying, 
\begin{align}
    2 - 2g = \frac{1}{2\pi}\int_{\mathcal{A}_g} d^{\,2}r_{ij}\,K\exp(2\Phi(\vec{r}_{ij})),
\end{align}
\blue{can be} induced by external electric fields,
\begin{align}\label{Liouville_eq}
    \vec{E} \propto \vec{\nabla}\Phi = -K\exp(2\Phi),
\end{align}
which create or annihilate vacancies from the lattices should de-intercalation or intercalation processes in the bilayered material be electrochemically permitted.\cite{kanyolo2022cationic, kanyolo2022advances2, liu2007} One can thus conclude that the non-trivial solutions for $\zeta(s) = 0$ corresponding to mass-less \blue{particles} on the honeycomb lattice should correspond to actual magnetic field terms applied to the bilayered material. The Riemann hypothesis\cite{conrey2003riemann} asserts that $s = 1/2 + i\gamma$, where our framework requires the so-called essential zeros at $\gamma = \gamma_g$ correspond to the flux values,
\begin{align}\label{gamma_B_eq}
    \gamma_g = \frac{q}{2\pi}\int_{\mathcal{A}_g}d^{\,2}r_{ij}\,B(\vec{r}_{ij}),
\end{align}
due to the \blue{$z$ component of the} external magnetic field $B$ \blue{given by eq. (\ref{B_eq})}, where $q$ is the elementary/Cooper-pair charge. Thus, the mass term is modulated to vanish accordingly \blue{for} $\gamma_g \neq 0$\blue{. The first three positive and negative values have been plotted} in Figure \ref{Fig_5}. 

To \blue{reflect} these two pieces of actual and pseudo-magnetic information in a consistent mathematical framework, we propose a novel complex-valued tensor equation similar to the idealised model\cite{kanyolo2020idealised}, 
\begin{subequations}\label{Idealised_eq}
\begin{align}
    \partial_vK_{uv} = -4\pi\ell\Psi^*\partial_u\Psi,\\
    K_{uv} = \partial_u\partial_v\Phi + i\frac{2q}{\ell}\epsilon_{uvw}A_w,
\end{align}
\end{subequations}
where $\ell$ is a cut-off scale for electromagnetic interactions along the $z$ direction, \blue{$\partial_u$ are partial derivatives,} $K_{uv} = K_{vu}^*$ is a complex-hermitian tensor, the repeated Euclidean indices $u, v$ and $w$ are summed over, $\Psi = \sqrt{\rho}\exp(iS)$, $A_w$, $\Phi$, $\epsilon_{uvw}$ are the vacancy wave function, the electromagnetic vector potential, the Liouville field and the completely anti-symmetric Levi-Civita symbol respectively. One can check that, the vacancy number density in 3D is given by $\rho = -(K/2\pi \ell)\exp(2\Phi)$ and $S$ is the quantum phase.  

Using this equation, we can introduce new complex variables $s$, $\overline{s}$ simply as, 
\begin{subequations}\label{s_calculation_eq}
\begin{multline}
    \hat{s} = \ell\int_0^{\ell}\langle \Psi(z)|\partial_{z}|\Psi(z) \rangle_g dz 
    = \ell\int_0^{\ell}\int_{\mathcal{A}_g}\Psi^*(\vec{r}_{ij}, z)\partial_z\Psi(\vec{r}_{ij}, z)d^{\,2}r_{ij}dz \\
    = \frac{\ell}{2}\int_{\mathcal{A}_g}\left (|\Psi(\vec{r}_{ij}, \ell)|^2 - |\Psi(\vec{r}_{ij}, 0)|^2\right )d^{\,2}r_{ij} 
    + i\ell\int_0^{\ell}\int_{\mathcal{A}_g}\partial_zS(\vec{r}_{ij}, z)|\Psi(\vec{r}_{ij}, z)|^2d^{\,2}r_{ij}dz\\
    = -g + i\ell \langle \partial_z S\rangle_g = -g + i\gamma_g,
\end{multline}
and, 
\begin{multline}
    \hat{s}^* = \ell\int_0^{\ell}\langle \Psi^*(z)|\partial_{z}|\Psi^*(z) \rangle_g dz 
    = \ell\int_0^{\ell}\int_{\mathcal{A}_g}\Psi(\vec{r}_{ij}, z)\partial_z\Psi^*(\vec{r}_{ij}, z)d^{\,2}r_{ij}dz\\
    = \frac{\ell}{2}\int_{\mathcal{A}_g}\left (|\Psi(\vec{r}_{ij}, \ell)|^2 - |\Psi(\vec{r}_{ij}, 0)|^2\right )d^{\,2}r_{ij} 
    - i\ell\int_0^{\ell}\int_{\mathcal{A}_g}\partial_zS(\vec{r}_{ij}, z)|\Psi(\vec{r}_{ij}, z)|^2d^{\,2}r_{ij}dz \\
    = -g - i\ell \langle \partial_z S\rangle_g = -g - i\gamma_g,
\end{multline}
\end{subequations}
where $\vec{r}_{ij} \in R^2$ and we have defined $\ell\int_{\mathcal{A}_g}|\Psi(\vec{r}_{ij}, 0)|^2d^{\,2}r_{ij} = 2(g - 1)$ for $g \neq 0$ and $\ell\int_{\mathcal{A}_g}|\Psi(\vec{r}_{ij}, \ell)|^2d^{\,2}r_{ij} = -2$ corresponding to the negative values of the Euler characteristic, $-E(\mathcal{A}_g) = 2g - 2$ before ($g \neq 0$) and after ($g = 0$) bifurcation respectively, and $\langle \partial_zS \rangle_g  = \int_0^{\ell}\int_{\mathcal{A}_g} \partial_zS(\vec{r}_{ij}, z) |\Psi(\vec{r}_{ij}, z)|^2d^{\,2}r_{ij}dz$. 

Thus, from eq. (\ref{Idealised_eq}), it is clear that \blue{a finite genus due to the external electric field in eq. (\ref{Liouville_eq}) corresponds to the sum},
\begin{align}
    s = \hat{s} + \hat{s}^* = -2g,
\end{align}
\blue{whereas} a finite magnetic flux due to the external magnetic field \blue{in eq. (\ref{gamma_B_eq})} corresponds to the imaginary part of $\hat{s}$ \blue{or} $\hat{s}^*$ \blue{depending on the direction of $B$} commensurate with eq. (\ref{zeta_eq}). Moreover, setting $S(\vec{r}_{ij}, z) + \Phi(\vec{r}_{ij}, z) = 0$ in eq. (\ref{s_calculation_eq}) guarantees flux is quantised $\gamma_g = g$, since there will be no distinctions between pseudo-magnetic and magnetic degrees of freedom, 
\begin{align}
    |\Psi(\vec{r}_{ij}, z)|^2 = -K\exp(2\Phi(\vec{r}_{ij}, z))/2\pi\ell = \frac{B(\vec{r}_{ij})}{4\pi\ell}.
\end{align}
For instance, for 1D systems in a ring exhibiting Peierls instability, there are no distinctions between the pseudo-magnetic and actual magnetic degrees of freedom, which should lead to a magnetisation effect proportional to the order parameter\cite{liang2006peierls}. However, we are interested instead in a different scenario where the actual magnetic and pseudo-magnetic effects are distinguishable, \textit{i.e.} $S(\vec{r}_{ij}, z)$ and $\Phi(\vec{r}_{ij}, z)$ play complementary \textit{albeit} antagonistic role in the realisation of a stable bilayered structure. An important scheme that implements this is to introduce quantum operators/averaging on $g = \nu - 1$ treated as quantum harmonic oscillator operators.\cite{kanyolo2020idealised} Setting $g = aa^{\dagger}$, $\nu = a^{\dagger}a$ where $a^{\dagger}, a$ are the creation and annihilation operators satisfying the bosonic commutation relation $[a^{\dagger}, a] = 1$, we first take the quantum average of the genus via a weighted sum, 
\begin{subequations}
\begin{align}
    \langle g \rangle_{\mathcal{P}} = \sum_{\nu = 0}^{\nu = \infty}g(\nu)\mathcal{P}_{\nu}(\beta) = \frac{1}{1 - \exp(-\beta)},
\end{align}
where $H_{\nu} = \nu + 1/2 = g - 1/2$ is the appropriately normalised Hamiltonian,
\begin{align}
    \mathcal{P}_{\nu}(\beta) = \frac{\exp(-\beta H_{\nu})}{\sum_{\nu = 0}^{\nu = \infty}\exp(-\beta H_{\nu})},
\end{align}
is the probability for a certain genus to occur and $\beta$ plays the role of the inverse temperature. Proceeding, using the gamma function, $\Gamma(\hat{s} = -2g)$ as a distribution function in a subsequent averaging\cite{kanyolo2022advances2} yields the result we seek, 
\begin{align}
     \langle\,\langle g \rangle_{\mathcal{P}}\,\rangle_{\Gamma (-2g)} = \frac{1}{\Gamma(-2g)}\int_0^{\infty} \frac{d\beta}{\beta}\beta^{-2g} \langle g \rangle_{\mathcal{P}}\exp(-\beta)\\
    = \frac{1}{\Gamma(-2g)}\int_0^{\infty} \frac{d\beta}{\beta}\beta^{-2g}\frac{1}{\exp(\beta) - 1} = \zeta(-2g), 
\end{align}
\end{subequations}
subject to zeta function regularisation by eq. (\ref{analytic_eq}) for values ${\rm Re}(s) \blue{= \delta} < 1$. Evidently, 
\begin{subequations}\label{tilde_s_eqs}
\begin{align}
    \tilde{s} = \langle \langle \hat{s} \rangle_{\mathcal{P}} \rangle_{\Gamma(-2g)} = \Delta(d) + i\gamma_g,\\
    \tilde{s}^* = \langle \langle \hat{s}^* \rangle_{\mathcal{P}} \rangle_{\Gamma(-2g)} = \Delta(d) - i\gamma_g,
\end{align}
\end{subequations}
where the fluxes $\gamma_{g \neq 0} \neq 0$ must satisfy $\zeta(s) = 0$ in order to guarantee the vanishing of the interaction $U_{R^2}(\tilde{s}) = U_{R^2}(\tilde{s}^*) = 0$, whose functional form is defined in eq. (\ref{continuum_h_Eq2}). This means that, for tuning of the mass term to its critical value by an external field to be possible, the fluxes in each hexagonal lattice ($hx, hx^*$) ought to occur at the essential zeros of $\zeta(s)$ corresponding to ${\rm Re}(s) = \Delta(d = 3) = 1/2$.

\begin{figure}
\begin{center}
\includegraphics[width=\columnwidth,clip=true]{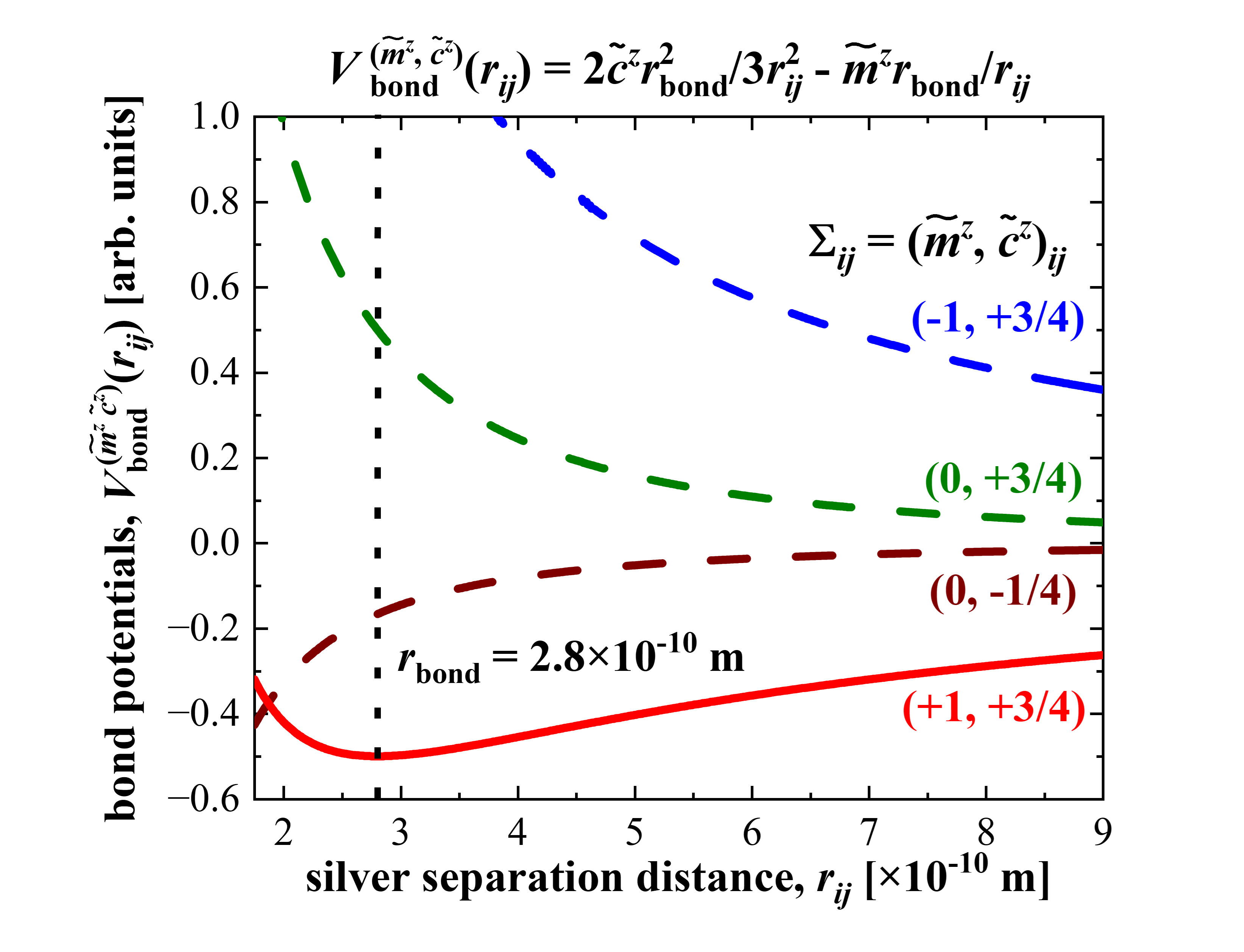}
\caption{Plot of the bond potentials $V_{\rm bond}^{(\tilde{m}^z, \tilde{c}^z)}(r_{ij})$ appropriately defined in eq. (\ref{Bonding_eq}), where $\tilde{m}^z$ and $\tilde{c}^z$ correspond to pseudo-magnetisation and correlation values of the wave functions $\Sigma_{ij}$ in eq. (\ref{pseudo_H_eq}). It is clear that $\Sigma_{ij} = (+1, +3/4)$ represents the silver-silver bond potential configuration with \blue{the argentophilic bond set at} $r_{\rm bond} = 2.8$ \AA.}
\label{Fig_6}
\end{center}
\end{figure}

Finally, the argentophilic bonds described by eq. (\ref{Heisenberg_eq1}) and \blue{hence} eq. (\ref{R3_eq}) in the \blue{two-silver} approximation \blue{ought to satisfy},
\begin{align}\label{Heisenberg_eq2}
    0 = 4J(\vec{r}_{ij})S_i^zS_j^z + \langle \langle h(\vec{r}_{ij}) \rangle \rangle (S_i^z + S_j^z),
\end{align}
where the \blue{double-averaging performed on $h(\vec{r}_{ij})$ given by} eq. (\ref{pseudo_h_eq}) \blue{corresponds to replacing $\hat{s}$ and $\hat{s}^*$ with their counterparts $\tilde{s}$ and $\tilde{s}^*$ respectively to yield}, 
\begin{align}
    \langle \langle h(\vec{r}_{ij}) \rangle \rangle = -\frac{m}{C}\left (\frac{\sqrt{2}\,a}{r_{ij}} \right )^{2(\tilde{s} + \tilde{s}^*)} = -\frac{Gm}{r_{ij}^2},
\end{align}
with $\vec{r}_{ij} \in R^3$ \blue{strictly in $R^3$}, $\tilde{s} + \tilde{s}^* = 1$ from eq. (\ref{tilde_s_eqs}), $G = 2a^2/C$ the square of the lattice constant with dimensions of [length]$^2$ and $J(\vec{r}_{ij}) = -A^2\chi_{\Delta(d)}(\vec{r}_{ij})$ the so-called Ruderman-Kittel-Kasuya-Yosida (RKKY) exchange interaction term in $d$ dimensions (Fourier transform of the Lindhard function in $d$ dimensions) characteristic of spin-orbit scattering of conduction electrons by the nonmagnetic \blue{(herein, pseudo-spin)} impurities in calculations involving $sd$ hybridisation\cite{aristov1997indirect} 
\blue{and},
\begin{subequations}
\begin{align}
    \chi_{\Delta}(\vec{r}_{ij})  = -\frac{1}{4\pi}\left (\frac{m_{\Delta}k_{\rm F}^2}{2\Delta + 1}\right )\left(\frac{k_{\rm F}}{2\pi r_{ij}}\right )^{2\Delta}f_{\Delta}(k_{\rm F}r_{ij}),\\
    f_{\Delta}(x) = \mathcal{J}_{\Delta(d)}(x)\mathcal{Y}_{\Delta(d)}(x) + \mathcal{J}_{\Delta(d) + 1}(x)\mathcal{Y}_{\Delta(d) + 1}(x).
\end{align}
\end{subequations}
Here, $m_{\Delta(d)}$ is the generated electron mass on the honeycomb lattice due to bifurcation, $A^2$ a proportionality constant and $\mathcal{J}_{\alpha}(x), \mathcal{Y}_{\alpha}(x)$ are Bessel functions of the first, second kind respectively. Since $\chi_{\Delta}(\vec{r}_{ij})$ \blue{is} proportional to the generated mass, $m_{\Delta(d)} = 2m\Delta(d)$, evidently $J(\vec{r}_{ij})$ is finite only in $d = 3$, allowing its \blue{positive} value to counteract the $\vec{r}_{ij} \notin R^2$ pseudo-spin term in eq. (\ref{R3_eq}) thus obtaining $H = 3J(\vec{r}_{ij}) + \langle \langle h(\vec{r}_{ij}) \rangle \rangle = 0$ \blue{as the final result of the calculations that commenced from} eq. (\ref{Heisenberg_eq1}). In $d = 3$ \blue{dimensions (3D)}, the RKKY terms are finite, 
\begin{align}
    \chi_{\Delta(3)}(r_{ij}) = \frac{mk_{\rm F}}{8\pi^3r_{ij}^3}\left (\frac{\sin\left (2k_{\rm F}r_{ij}\right )}{2k_{\rm F}r_{ij}} - \cos\left (2k_{\rm F}r_{ij}\right ) \right )
    = -mk_{\rm F}/8\pi^3r_{ij}^3,
\end{align}
where the last line is obtained by taking $\vec{r}_{ij}\cdot\vec{k}_{ij} = \pi N \neq 0$ and $\vec{r}_{ij}\times\vec{k}_{ij} = \vec{0}$. Here, $N \in \mathbb{N}$ is a non-vanishing positive integer, $\vec{r}_{ij} \neq 0, \vec{k}_{ij} \neq 0$ are both vectors in $R^3$, $|\vec{r}_{ij}| = r_{ij}$ and $|\vec{k}_{ij}| = k_{\rm F}$ defines the Fermi surface. 

The final expression for the argentophilic bond becomes, 
\begin{subequations}
\begin{align}\label{orbits_eq}
    0 = \frac{L^2}{m^2r_{ij}^3} - \frac{Gm}{r_{ij}^2} \equiv -\frac{\partial V(\vec{r}_{ij})}{\partial r_{ij}},
\end{align}
where,
\begin{align}
    V(\vec{r}_{ij}) = \frac{1}{2m^2}\frac{L^2}{r_{ij}^2} - \frac{Gm}{r_{ij}},
\end{align}
\end{subequations}
is the bond potential and we have introduced $A^2 \equiv 8\pi^3L^2/3k_{\rm F}m^3$ in order to write the interaction in a familiar form, analogous to gravitational orbits under the potential $V(\vec{r}_{ij})$ representing the emergent attractive forces experienced by conduction electrons \blue{and diffusing silver cations in the material} under `Newton's inverse-square law' with $G$ playing the role of a `gravitational constant' and $L$ the angular momentum\blue{,} leading to a conductor-semiconductor-insulator phase transition \blue{for all charge transport on the bifurcated lattice}. \blue{T}he argentophilic bond length \blue{corresponds to} $r_{\rm bond} = L^2/Gm^3 \leq 2.83$ \AA, \blue{which} \blue{displays} the relation between the constants $L$, $m$ and $G$ in terms of the observed bond length. However, since these constants are presently undetermined, for illustration purposes we shall set $r_{\rm bond} \equiv 2.8$ \AA \,\, and plot in Figure \ref{Fig_6} the appropriately normalised potential,
\begin{align}\label{Bonding_eq}
    V_{\rm bond}^{(\tilde{m}^z, \tilde{c}^z)}(r_{ij}) = 2\tilde{c}^zr_{\rm bond}^2/3r_{ij}^2 - \tilde{m}^zr_{\rm bond}/r_{ij},
\end{align}
for various values of $\Sigma_{ij} = (\tilde{m}^z, \tilde{c})$ in eq. (\ref{pseudo_eq}), thus illustrating the presence of a minimum at $r_{ij} = r_{\rm bond}$ only for the pseudo-spin wave function $\Sigma_{ij} = (+1, +3/4)_{ij}$ given in eq. (\ref{pseudo_eq}). Changing the sign of the pseudo-magnetic field $h(\vec{r}_{ij}) \blue{\rightarrow -h(\vec{r}_{ij})}$ \blue{exchanges pseudo-spin up with pseudo-spin down in the calculation in eq. (\ref{R3_eq}), thus selecting} instead $\Sigma_{ij} = (-1, +3/4)_{ij}$ as the potential with the minimum at $r_{ij} = r_{\rm bond}$.

\section{Discussion}

\blue{\textbf{Conductivity and localisation.}\textemdash 
Much like elemental silver, metallic bonds in monolayered silver-based honeycomb layered oxides are expected to exhibit higher electronic and ionic conductivity relative to alkali metal-based honeycomb layered oxides.\cite{kanyolo2021honeycomb} However, an energy/mass gap in the ionic/electronic density of states leads to a reduction in conductivity relative to the monolayered structure, hence the monolayer-bilayer phase transition can be interpreted as a conductor-semi-conductor-insulator phase transition. For instance, the honeycomb layered $\rm Ag_2Zn_2TeO_6$, which predominantly features domains with a monolayered arrangement of Ag atoms, demonstrates a conductivity (ionic + electronic) on the order of $\rm 10^{-3}$ [S/cm] at room temperature\cite{masese2022honeycomb}, which is orders of magnitude higher than \textit{e.g.} the conductivity of the monolayered $\rm K_2Mg_2TeO_6$ which exhibits ionic conductivity of order $\rm 10^{-6}$ [S/cm].\cite{kanyolo2021honeycomb} On the other hand, in honeycomb layered tellurates with a global composition of ${\rm Ag}_2^{1+}M_2\rm TeO_6$, ($M = \rm Ni, Co, Mg, Zn$ \textit{etc}), which manifest mixed \red{monolayered and} bilayered silver atom domains
, the conductivity is expected to be considerably lower than \red{the} pure monolayered phase due to the presence of bilayered domains such as $\rm Ag_6Mg_2TeO_6$. In fact, the reported ionic conductivity of, 
\textit{e.g.} the global composition $\rm Ag_2Mg_2TeO_6$ is on the order of $\rm 10^{-6}$ [S/cm] at room temperature. \cite{masese2022honeycomb} Specifically, `global composition' means that the monolayered ${\rm Ag}_2^{1+}M_2\rm TeO_6$ phase (expected to contribute to the bulk of the conductivity) is a mixed phase, characterised by bifurcated domains corresponding to the bilayered ${\rm Ag}_6^{v+}M_2\rm TeO_6$ whose silver ions are experimentally and theoretically determined to be subvalent (experimentally: $1/2+ \leq v \leq 2/3+$; theoretically: $v = 1/2+$). Whilst we cannot presently analytically estimate these electronic and ionic conductivity values, heuristically, 
any conductivity ($\sigma$) 
\red{derivation} assuming locally gapped hopping events between two Ag atoms on the (bifurcated) honeycomb lattice \red{would follow Fermi's golden rule}, thus resembling \red{typical} calculations 
\red{of Cooper-pair tunnelling}\cite{kanyolo2021renormalization}, 
\begin{align}
    \sigma (gap) \propto \sum_{particle\,\,momenta}\times \sum_{hole\,\,momenta} \rightarrow  \int d(particle\,\,energy)\times \int d(hole\,\,energy).
\end{align}
Meanwhile, the ratio of conductance (ratio of density of states) is expected to roughly scale as the Jacobian\cite{kanyolo2021renormalization},
\begin{align}\label{Jacobian_eq}
  \left |\frac{\sigma(U_{R^2})}{\sigma(0)}\right | \propto \frac{\int d(quasiparticle\,\,energy)d(quasihole\,\,energy)}{\int d(particle\,\,energy) d(hole\,\,energy)} \propto \left (\frac{d\mathcal{E}}{dE} \right )^2 = \frac{E^2}{E^2 + U_{\rm R^2}^2} \equiv \mathcal{J},
\end{align}
where $0 \leq \mathcal{J} \leq 1$ is the Jacobian, $\sigma(U_{R^2})$ and $\sigma(0) \neq 0$ are the conductivity of the bifurcated ($U_{R^2}^2 \neq 0$) and non-bifurcated ($U_{R^2}^2 = 0$) honeycomb lattices respectively 
\red{along} the $a-b$ plane of the material, $\mathcal{E} = \sqrt{E^2 + U_{R^2}^2}$ is the quasi-particle energy, $E = |\vec{p}|$ is the particle/hole momentum and $U_{R^2}$ is the energy/mass gap introduced by the bifurcation. Complete localisation of the particle implies that $\vec{p} = 0$, which is 
\red{precluded} by Lorentz invariance 
\red{(massless particles in a relativistic theory cannot be stationary, \textit{i.e.}} $\vec{p} = \vec{k}_{\rm F}$, where $k_{\rm F}$ is the Fermi energy) in 2D and thermal agitation $\langle p^2 \rangle = 3|U_{\rm R^2}|k_{\rm B}T/2 \neq 0$ in 3D with $k_{\rm B}$ Boltzmann constant and $T$ the \red{equilibrium} temperature. Thus, the mass/energy gap determines the conduction properties of the Ag bilayer $\sigma(U_{R^2}) = 0$ whereby $U_{R^2} \neq 0$ is a semi-conductor and $U_{R^2} = 0$ is a metal.} \\

\blue{ \textbf{Electron localisation function paradigm.}\textemdash Electron Localisation Function (ELF) analysis is certainly a potent tool 
\red{for} determining whether electron localisation occurs. However, the major challenges to successfully justifying the results of ELF analysis for Ag bilayered structures exhibiting both argentophilicity and subvalency primarily lie in establishing a suitable rationale behind the particular bonding theory implemented in \textit{ab initial} DFT simulation packages.\cite{hafner2008ab, erba2022crystal23, sangalli2019many} Within the ELF paradigm\cite{savin1997elf}, given the core and/or valence electron(s) eigenfunction(s) $\varphi_i^\sigma$ of spin $\sigma$ obtained by solving the Schrodinger equation $i\hbar\partial \varphi_i^{\sigma}/\partial t = H\varphi_i^{\sigma}$ either by perturbation theory (e.g. Hatree-Fock) or diagonalising the Hamiltonian, $H$ (e.g. Kohn-Sham formalism), the nature of the Hamiltonian, $H = -\red{\nabla^2}/2m_{\rm e} + U$ ($\red{\nabla^2}$ is the Laplacian, $m_{\rm e}$ is the electron mass) is governed by the kinetic energy $\red{-\nabla^2}/2m_{\rm e}$ alongside the particular (pseudo-)potential(s), $U$ employed, which in turn determine the eigenfunctions $\varphi_i^{\sigma}$. 

Since we are interested in the behaviour of valence/conduction electrons with differing chirality, we can define their spin density as the index of two pseudo-spin quantities,
\begin{align}\label{index_eq}
    \rho_{\sigma} = \sum_{i = 1}^{N}|\varphi_i^{\sigma}|^2 - \sum_{i = 1}^{N}|\varphi_i^{-\sigma}|^2 = \psi_{\sigma}^{\dagger}\gamma^0\psi_{\sigma} = \overline{\psi_{\sigma}}\psi_{\sigma},    
\end{align}
where $2N$ is the total number, $\psi_{\sigma} = (\varphi_1^{\sigma}, \varphi_1^{-\sigma}, \cdots, \varphi_N^{\sigma}, \varphi_N^{-\sigma})^{\rm T}$, $\gamma^0\psi_{\sigma} = (\varphi_1^{\sigma}, -\varphi_1^{-\sigma}, \cdots, \varphi_N^{\sigma}, -\varphi_N^{-\sigma})^{\rm T}$, \red{superscript-}$T$ is the transpose, $\psi_{\sigma}^{\dagger} = \psi^{*{\rm T}}$, $\overline{\psi_{\sigma}} = \psi_{\sigma}^{\dagger}(\gamma^0)^{-1}$ and $(\gamma^0)^{-1} = \red{\tau_3}\otimes I_N = \gamma^0$ with $\tau_3$ the $z$ component of the Pauli vector and $I_{\rm N}$ the $N\times N$ identity matrix. Meanwhile, the valence electron localisation measure 
\red{becomes},
\begin{align}\label{measure_eq}
    \mathcal{I}_{\sigma} = \overline{\psi_{\sigma}}\left (\nabla^2 - \left (\frac{1}{2}\vec{\nabla}\ln \rho_{\sigma}\right )^2\right )\psi_{\sigma}, 
\end{align}
where 
the Laplacian 
essentially acts to reproduce the sum of kinetic energy terms of the wave functions $\varphi_i^{\pm \sigma}$. Meanwhile, the negative definite potential energy term $-(\vec{\nabla}\ln \rho_{\sigma})^2/4$ negates the kinetic energy contribution of bosonic modes due to \textit{e.g.} pairing of valence electrons potentially leading to localisation. Thus, the 
ELF is defined by, 
\begin{align}
   0 \leq {\rm ELF} = \frac{1}{1 + (\mathcal{I}_{\sigma}/\mathcal{I}_{\sigma}^0)^2} \leq 1,
\end{align}
where $\mathcal{I}_{\sigma}^0$ is the corresponding localisation measure when the valence electrons in the computation is replaced by a uniform electron gas. Since determining eq. (\ref{index_eq}) is not only almost always computationally intensive \textit{e.g.} involving finding the eigenfunctions via the Kohn-Sham formalism or Hartree fock, the bottle-neck in determining ELF corresponds to whether the phenomena being described are amenable to these Hamiltonian diagonalisation methods. Unfortunately, DFT descriptions involving phase transitions such as (those) described by the BCS theory tend to be highly complex.\cite{linscheiddensity} Some of the present challenges inherent in incorporating the $SU(2)\times U(1)$ model\cite{masese2022honeycomb} within \textit{ab initio} DFT calculations have been reviewed in a recent review paper.\cite{kanyolo2023honeycomb} 

Nonetheless, the metal-insulator properties can be extracted from the many-body Green's function suitably treated, provided the simulation package is specifically tailored to simulate phase transitions.\cite{oliveira1988density, marques2005ab, linscheid2015ab} Since our treatments lead to localisation of valence electrons by 
\red{Cooper-}pairing, one should eventually be able to analytically link the pseudo-spin model to aspects of ELF analysis, thus potentially by-passing the explicit reliance of \textit{ab initio} DFT to obtain $\varphi_i^{\sigma}$ and $\rho_{\sigma}$ relevant for our pseudo-spin model. We find that, a great deal of simplification can be achieved by identifying ELF in terms of the order parameter/mass gap, $U_{R^2}$. In the pseudo-spin model, it certainly self-consistent to take, 
\begin{align}\label{Jacobian_ELF_eq}
    {\rm ELF} = 1 - \mathcal{J} = \frac{1}{1 + (E^2/U_{R^2}^2)},
\end{align}
where $\mathcal{J}$ is the Jacobian introduced in eq. (\ref{Jacobian_eq}), suggesting the localisation measure scales as the particle momentum $E = |\vec{p}|$ as expected and $U_{R^2}$ plays the role of the localisation measure of the uniform electron gas. For instance, using the coherence factors given in eq. (\ref{coherence_eq}) equivalent to $\mu_{\pm} = \sqrt{(1 \pm E/\mathcal{E})/2}$, we find eq. (\ref{Jacobian_ELF_eq}) is transformed into, 
\begin{align}\label{det_A_eq}
    \det
\begin{pmatrix}
d\mathcal{E}/dE & -2\mu_-^2\\ 
2\mu_+^2 & d\mathcal{E}/dE
\end{pmatrix} 
= \det(A) = 1,
\end{align}
with $\det(A)$ the determinant of the matrix $A$. Thus, while the two formalisms are essentially equivalent under certain assumptions\cite{kanyolo2023honeycomb, kanyolo2022advances2}, the pseudo-spin model certainly has advantages over the fully-fledged $SU(2)\times U(1)$ model since it appears more amenable to typical methods employed in \textit{ab initio} calculations.\cite{garcia1992dimerization, said1984nonempirical}}

\blue{\textbf{Pseudo-magnetic response of the rest of the material.}\textemdash It is evident that the pseudo-spin model treats the Ag honeycomb lattice as isolated from the rest of the material. This approximation is applied due to our emphasis on obtaining analytic results that comprehensively capture the essence of the monolayer-bilayer phase transition. Nonetheless, we wish to discuss the potential limits of this `spherical cow' approximation to the pseudo-spin model. First, the effect of (potentially) magnetic $M^{2+} = \rm Cr^{2+}, Mn^{2+}, Fe^{2+}, Co^{2+}, Ni^{2+}$ \textit{etc} 
\red{ions} in $M_2\rm TeO_6$ 
\red{has been} neglected in our treatment. This restricts the search for the proposed magnetic response of the Ag bilayered structure in ${\rm Ag}_6M_2\rm TeO_6$ to non-magnetic materials \red{(where} $M^{2+} = \rm Mg^{2+}, Zn^{2+}$ \textit{etc} is non-magnetic\red{)} for unambiguous results. \red{Second, i}n the case of the Ag bilayered materials ${\rm Ag}_2^{1/2+}M^{3+}\rm O_2^{2-}$ with $M^{3+} = \rm Cr^{3+}, Mn^{3+}, Fe^{3+}, Co^{3+}, Ni^{3+}$ \textit{etc}, provided there is a Jahn-Teller distortion in the $M^{3+}\rm O_2^{2-}$ layer, the most suitable candidate to test the magnetic response of the Ag bilayered structure within the pseudo-spin model is surprisingly $\rm Ag_2^{1/2+}Co^{3+}O_2^{2-}$\red{. This follows} since the energetically stable electronic configuration is expected to be given by the non-magnetic ${\rm Co}^{3+}\,\,t_{2g}^{6+}e_g^0$ with the effective spin $S = 0$.\cite{kanyolo2023honeycomb} 

Finally, we need to also consider the effects of the spontaneous pseudo-magnetisation displayed in Figure \ref{Fig_3} on the rest of the material. Restricting our attention to the $M_2\rm TeO_6$ layers of ${\rm Ag}_6M_2\rm TeO_6$, 
\red{we can employ} the FCC/HCP notation\cite{kanyolo2022advances2} 
\red{to denote} its stacking sequence 
as, 
\begin{align}\label{stacking_sequence_eq}
    U_{\rm O}V_{(M, M,{\rm Te})}W_{\rm O}W_{\rm Ag}V_{\rm Ag}V_{\rm O}U_{(M, M,{\rm Te})}W_{\rm O},
\end{align}
where $U, V, W$ are each hexagonal lattices and \red{their stacking} $UVW$ comprises a larger hexagonal lattice \red{as viewed along the $c$-zone axis}. For instance, eq. (\ref{stacking_sequence_eq}) indicates that Ag in ${\rm Ag}_6M_2\rm TeO_6$ has a dumbbell/linear coordination to O atoms. The slab corresponds to $U_{\rm O}V_{(M, M,{\rm Te})}W_{\rm O}$. Meanwhile, the stacking sequence of the experimentally stable Ag monolayered material ${\rm Ag}_2M_2\rm TeO_6$ is, 
\begin{align}\label{stacking_sequence_eq2}
    U_{\rm O}V_{(M, M,{\rm Te})}W_{\rm O}W_{\rm Ag}W_{\rm O}U_{(M, M,{\rm Te})}V_{\rm O}.
\end{align}
Thus, eq. (\ref{stacking_sequence_eq2}) is transformed into eq. (\ref{stacking_sequence_eq}) by an insertion of a single Ag hexagonal layer (\textit{i.e.} $V_{\rm Ag}$) and a shear transformation (implementing an inversion) of the $M_2\rm TeO_6$ slab (\textit{i.e.} $W_{\rm O}U_{(M, M,{\rm Te})}V_{\rm O} \rightarrow V_{\rm O}U_{(M, M,{\rm Te})}W_{\rm O}$). However, since the Ag monolayers in eq. (\ref{stacking_sequence_eq2}) are hexagonal (triangular, not honeycomb), the monolayer-bilayer phase transition is achieved by Ag intercalation which tends to saturate the material with Ag ions rather than the bifurcation mechanism introduced by the pseudo-spin model.\cite{kanyolo2022advances} 

Meanwhile, DFT computations (that did not take into account argentophilicity and Ag subvalency) predicted a silver-based honeycomb layered tellurate with a different Ag honeycomb monolayered phase\cite{tada2022implications}, 
\begin{align}\label{stacking_sequence_eq3}
    U_{\rm O}V_{(M, M,{\rm Te})}W_{\rm O}W_{(\rm Ag, Ag, - )}W_{\rm O}U_{(M, M,{\rm Te})}V_{\rm O},
\end{align}
which \red{presumably was difficult to realise} 
experimentally \red{without applying an external magnetic field}
\red{to suppress} finite argentophilic interactions that \red{would} result in bifurcation. Thus, our pseudo-spin model applies to this precursor to eq. (\ref{stacking_sequence_eq}) related to eq. (\ref{stacking_sequence_eq}) by bifurcation ($W_{(\rm Ag, Ag, - )} \rightarrow W_{(\rm Ag, - , - )}V_{(\rm Ag, -, -)}$), and \red{subsequent} saturation $W_{(\rm Ag, - , - )}V_{(\rm Ag, -, -)} \rightarrow W_{(\rm Ag, Ag, Ag)}V_{(\rm Ag, Ag, Ag)}$. Moreover, all these transformations preserve the rest of the structure ($M_2\rm TeO_6$ layers) \textit{modulo} the \red{four} Burgers vectors $[\pm 2/3, +1/3, 0]$ and $[\pm 2/3, -1/3, 0]$, implemented by \red{the} shear transformations\cite{masese2021mixed},
\begin{align}
    S_+ = \begin{pmatrix}
1 & 0 & \pm 2/3\\ 
0 & 1 & +1/3\\ 
0 & 0 & 1
\end{pmatrix},\,\,\,S_- = \begin{pmatrix}
1 & 0 & \pm 2/3\\ 
0 & 1 & -1/3\\ 
0 & 0 & 1
\end{pmatrix},
\end{align}
respectively.}\\

\blue{
\textbf{The Ag subvalent state.}\textemdash $SU(2)\times U(1)$ model predicts that the 
\red{Ag subvalent} state of $1/2+$ is a massive valence state of silver, where 
\red{the finite mass leads to} a bifurcated honeycomb lattice structure in honeycomb layered frameworks with the chemical formula written as $\rm Ag_2^{1/2+} = Ag^{2+}Ag^{1-}$. Whilst this notation is useful, it can be misleading since it appears to suggest that the \red{individual} $\rm Ag^{2+}$ and $\rm Ag^{1-}$ valency states can be resolved experimentally. This would also erroneously predict that 
\red{the $1/2+$ and $2/3+$ subvalent states} would 
obviously \red{be} magnetic by virtue of $\rm Ag^{2+}$ 
\red{with} electronic state 
$4d^95s^0$), whilst the electronic states of $\rm Ag^{1+}$ and $\rm Ag^{1-}$ are $4d^{10}5s^0$ and $4d^{10}5s^2$ respectively. However, 
\red{analogous to the fact} that the electron and neutrino of the standard model are \textit{bona fide} particles in electroweak \red{theory of the standard model of particle physics},  according to the $SU(2)\times U(1)$ model\cite{masese2022honeycomb, kanyolo2023honeycomb}, the Ag subvalent states of $1/2+$ and $2/3+$ observed in silver-based bilayered structures are suggested to arise from a quantum superposition of $\rm Ag^{1+}, Ag^{2+}$ and $\rm Ag^{1-}$ valency states easily represented in Bloch sphere coordinates,
\begin{multline}\label{hybridisation_eq}
    |\theta, \varphi \rangle = \cos(\theta)|{\rm Ag^{1+}}\rangle + \sin(\theta)\cos(\varphi)|{\rm Ag^{1-}}\rangle + \sin(\theta)\sin(\varphi)|{\rm Ag^{2+}}\rangle\\
    = \cos(\theta)|4d_{z^2}^2\rangle |5s^0 \rangle + \sin(\theta)\cos(\varphi)|4d_{z^2}^2\rangle |5s^2 \rangle + \sin(\theta)\sin(\varphi)|4d_{z^2}^1\rangle |5s^0 \rangle, 
\end{multline}
where $\varphi$ and $\theta$ are the azimuthal and polar angles respectively. This crucially implies that the $4d^95s^0$ electronic configuration responsible for $\rm Ag^{2+}$ state does not occur in isolation when Ag is subvalent.

To be consistent with $SU(2)\times U(1)$ \red{model\cite{masese2022honeycomb}}, the angles can only take on particular discrete values, \textit{i.e.} $\theta = 0, \pi/4, \pi/2$ and $\varphi = 0, \pi/4, \pi/2$. Within this picture, $|0, \varphi \rangle = |\rm Ag^{1+} \rangle$, $|0, \pi/2\rangle = |\rm Ag^{1-} \rangle$ and $|\pi/2, \pi/2 \rangle = |\rm Ag^{2+} \rangle$ form a Hilbert space. The fact that a vacuum state constructed from $\rm Ag^{1+}$ admits a continuum of $\varphi$ values is indicative of extreme degeneracy under $sd$ hybridisation, which makes its vacuum unstable. Crucially, we can consider a new basis of the Hilbert space, whereby the valence states $|\pi/4, 0\rangle = |4d_{z^2}^2\rangle (|5s^0 \rangle + |5s^2 \rangle)/\sqrt{2}$ and $|\pi/4, \pi/2\rangle = |5s^0\rangle (|4d_{z^2}^2 \rangle + |4d_{z^2}^1\rangle)/\sqrt{2}$ are separable and hence not entangled, whereas $|{\rm Ag}^{1/2+} \rangle = |\pi/2, \pi/4\rangle = (|4d_{z^2}^2\rangle |5s^2 \rangle + |4d_{z^2}^1\rangle |5s^0 \rangle)/\sqrt{2}$ is inseparable, and hence an entangled state. Thus, while useful within a chemical formula, the notation $\rm Ag_2^{1/2+} = \rm Ag^{2+}Ag^{1-}$ is indeed misleading since it presupposes $|{\rm Ag_2^{1/2+}}\rangle = |{\rm Ag^{2+}}\rangle|{\rm \red{Ag^{1-}}}\rangle$ which is evidently magnetic due to the involvement of the separable $\rm Ag^{2+}$ state. In other words, no such guarantee exists for the entangled state, $\rm Ag^{1/2+}$ which should be treated as a \textit{bona fide} valency state. However, it appears $\rm Ag_2^{1/2+}Ag^{1+}$ (\textit{e.g.} in $\rm Ag_2^{1/2+}Ag^{1+}Ni_2^{3+}O_4^{2-}$) is a valid hybrid valency state arising from alternating Ag monolayers and bilayers and hence \red{the $2/3+$ subvalent state} is likely not a \textit{bona fide} valency state such as $\rm Ag_2^{1/2+}$.\cite{masese2022honeycomb} Given the novelty of the $SU(2)\times U(1)$ model to describing Ag bilayered structures\cite{masese2022honeycomb}, the precise (pseudo-)magnetic nature of the proposed $\rm Ag_2^{1/2+}$ state 
\red{remains} unverified, although Ag bilayered materials exhibit a relatively weak unexplained \red{(anti-)}ferromagnetism.\cite{taniguchi2020butterfly, komori2023antiferromagnetic} Finally, \red{we note that} the pseudo-spin model implements the different concept of pseudo-spin (and pseudo-magnetic fields) which arises from the bipartite nature of the honeycomb lattice via two orthogonal chiral states, 
\begin{align}
    |{\rm Ag}^{1/2+}\rangle_\pm = |\pi/2, \pm \pi/4 \rangle = (|4d_{z^2}^2\rangle |5s^2 \rangle \pm |4d_{z^2}^1\rangle |5s^0 \rangle)/\sqrt{2}.
\end{align}
For relativistic massless particles, there is no distinction between helicity and chirality, meaning the conduction electron spins constitute the chiral degrees of freedom. Heuristically, $sd$ hybridisation identifies the $d$-band with the conduction band and $s$-band with the valence band, thus relating this chirality/pseudo-spin on the honeycomb lattice with the spin of the $d_{z^2}^1$ electron, serving as the premise of our pseudo-spin model.}

\textbf{Conclusion.}\textemdash Exploiting the pseudo-spin degree of freedom availed by the bipartite honeycomb lattice of Ag cations inherited from the spin of their half-filled $4d_{z^2}$ orbitals, we introduced a pseudo-spin Heisenberg Hamiltonian to describe the argentophilic bond observed in silver-based bilayered materials, arriving at the same qualitative results as a previously proposed $SU(2)\times U(1)$ model.\cite{masese2022honeycomb} \blue{For instance, both models predict a spontaneously generated Dirac mass for the Ag cations and valence electrons on the honeycomb lattice, responsible for argentophilicity and localisation respectively. Since the spontaneously generated valence electron mass likely differs from the actual electron mass, this `renormalisation' mechanism potentially explains the heavy fermion behaviour reported in several silver-based honeycomb layered materials whenever $m \gg m_{\rm e}$ with $m$ the generated Dirac mass of the valence electron and $m_{\rm e}$ the mass of the electron.\cite{eguchi2010resonant, ji2010orbital, kanyolo2023honeycomb}} In the Heisenberg Hamiltonian of the pseudo-spin model, the monolayer-bilayer phase transition occurs due to the spontaneous emergence of a pseudo-magnetic interaction term, with the mechanism for Ag bilayers reminiscent of Zeeman splitting. The advantage conferred by the proposed approach is the novel possibility of engineering a crossover between 2D and 3D behaviour of \blue{silver ions} and their valence electrons expected to be responsible for \blue{particle pairing/bonding} and a metal-semiconductor-insulator phase transition. Incidentally, since excellent conductors such as elemental silver are expected to be poor BCS superconductors due to fairly weak electron-phonon coupling\cite{tinkham2004introduction}, the electron pairing mechanism herein incidentally explains the low temperature superconductivity reported in $\rm Ag_2^{1/2+}F$, \textit{albeit} \blue{remains} a normal conductor instead of semi-conductor or insulator\cite{wang1991anisotropic} at temperatures well-above the reported transition temperature ($T_{\rm c} \simeq 66$ mK).\cite{andres1966superconductivity} Moreover, the silver bilayer in such materials is expected to be tuned not only by pseudo-magnetic fields in the form of stress and strain, but more readily by external magnetic fields whose flux values occur at the essential zeros of the so-called $L(s)$ functions such as the famous Riemann zeta function, corresponding to the Mellin transform of specific lattice theta functions.\cite{conrey2003riemann, kanyolo2022advances2} Such experiments should be performed on (near-)single crystals of pure silver-bilayered materials such as $\rm Ag_2^{1/2+}F$ or the more stable $\rm Ag_6^{1/2+}Mg_2TeO_6$ (or preferably any other silver bilayered material with non-magnetic slab cations in order to avoid effects such as magneto-resistance\cite{taniguchi2020butterfly} due to magnetic ordering of $M = \rm Ni, Co$ \textit{etc} cations in the slabs) prepared\blue{, for instance,} by mechanical exfoliation.\cite{taniguchi2020butterfly}
\blue{Finally, manifesting such a monolayer-bilayer phase transition experimentally would offer a potent tool to characterise and/or isolate factors related to the $M_2\rm TeO_6$ inter-layer interactions that potentially hinder the realisation of Kitaev physics in materials such as ${\rm Ag}_6M_2\rm TeO_6$ ($M = \rm Co, Ni$, {\textit etc}).\cite{komori2023antiferromagnetic} The value of this work also lies in the novel mathematical physics tools and ideas (related to emergent gravity in honeycomb layered materials and the zeta function\cite{kanyolo2022advances2, kanyolo2020idealised, kanyolo2022cationic}) employed in order to achieve the dual treatment of magnetic and pseudo-magnetic degrees of freedom on the honeycomb lattice.}\\

\textbf{Acknowledgements.}\textemdash This 
\blue{work was} supported in part by the AIST Edge Runners Funding and Iketani Science and Technology Foundation. \\

\textbf{Data availability.}\textemdash Data will be made available on request.

\bibliography{pseudo-spin}

\end{document}